\documentclass[a4paper]{article}
\usepackage{amsmath,amssymb,amsthm,amscd}
\def\be{\begin{equation}}
\DeclareMathOperator{\Ad}{Ad}

\DeclareMathOperator{\tr}{tr}
\DeclareMathOperator{\Diff}{Diff}
\DeclareMathOperator{\lspan}{span}
\DeclareMathOperator{\Hom}{Hom}

\DeclareMathOperator{\id}{id} \DeclareMathOperator{\im}{Im}
 
\DeclareMathOperator{\SO}{SO} 
 \DeclareMathOperator{\diag}{diag}
\DeclareMathOperator{\GL}{GL} 

\DeclareMathOperator{\RE}{Re}

\DeclareMathOperator{\const}{const}

\DeclareMathOperator{\Dom}{Dom}\DeclareMathOperator{\dist}{dist}
\newcommand{\ii}{\mathbf{i}}

\DeclareMathOperator{\loc}{loc}

\def\pd#1#2{\frac{\partial{#1}}{\partial{#2}}}
\def\pd1#1{\frac{\partial}{\partial#1}}
\def\d1#1{\frac{d}{d#1}}

\newcommand{\laplace}{\bigtriangleup}

\def\text#1{\mbox{#1}}

\newtheorem{theore}{Theorem}\newtheorem{Lem}{Lemma}
\newtheorem{Rem}{Remark}

\newtheorem{proposit}{Proposition}

\allowdisplaybreaks
\begin{document}

\author{Alexey V.~Shchepetilov\footnote{Department of Physics, Moscow State
University, 119992 Moscow, Russia, e-mail address:
quant@phys.msu.su}}
\title{Two-body quantum mechanical problem on spheres}
\date{}\maketitle
\begin{abstract}
The quantum mechanical two-body problem with a central interaction
on the sphere ${\bf S}^{n}$ is considered. Using recent results in
representation theory an ordinary differential equation for some
energy levels is found. For several interactive potentials these
energy levels are calculated in explicit form.

\vskip 20pt

\noindent PACS numbers: 03.65.Fd, 02.40.Vh, 02.40.Ky.\\
Mathematical Subject Classification: 43A85, 22E70, 57S25, 70G65.
\end{abstract}

\section{Introduction}\label{Introduction}\markright{\ref{Introduction}
Introduction}

The history of mechanics on constant curvature spaces started more
than one and a half century ago.

The analog of Newton (or Coulomb) force for the hyperbolic space
$\mathbf{H}^{3}$ was proposed already by founders of the
hyperbolic geometry N.I.~Lobachevski (in 1835-38) \cite{Lob} and
J.~Bolyai (between 1848 and 1851) \cite{Bol} as the value
$F(\rho)$, which is inverse to the area of the sphere in
$\mathbf{H}^{3}$ of radius $\rho$ with an attractive body in the
center.

The analytical expression for the Newtonian potential in the space
$\mathbf{H}^{3}$ was written in 1870 by E.~Schering \cite{Sch1}
(see also his paper \cite{Sch2} of 1873), without any motivation
and references to N.I.~Lobachevski and J.~Bolyai.

In 1873 R.~Lipschitz considered a one-body motion in a central
potential on the sphere $\mathbf{S}^{2}$ \cite{Lip3}. He knew that
the central potential $V_{c}$ satisfies the Laplace equation on
$\mathbf{S}^{3}$. However due to some reason he preferred to
consider another central potential $V(\rho)\sim\sin^{-1}(\rho/R)$,
where $\rho$ is a distance from the center and $R$ is a curvature
radius. He calculated the general solution of this problem through
elliptic functions.

In 1885 W.~Killing found the generalization of all three Kepler
laws for the sphere $\mathbf{S}^{3}$ \cite{Kil2}. He considered
the attractive force as an inverse area of a $2$-dimensional
sphere in $\mathbf{S}^{3}$ as N.I.~Lobachevski and J.~Bolyai did
before. In the next year these results was published also by
C.~Neumann in \cite{Neumann}. Their expansion onto the hyperbolic
case was carried out in H.~Liebman paper \cite{Lib02} in 1902 and
later in 1905 in his book on noneuclidean geometry \cite{Lib}.
Note that he started from ellipses in $\mathbf{S}^{3}$ or
$\mathbf{H}^{3}$ and derived a potential in such a way that the
first Kepler law would be valid. He derived also the
generalization of the oscillator potential for these spaces from
the requirement that a particle motion occurs along an ellipse
with its center coinciding with the center of the potential.

Also in the same paper \cite{Kil2} W.~Killing proved the variable
separation in the two-centre Kepler problem on the sphere
$\mathbf{S}^{n}$, which implies the integrability of this problem.

The well-known Bertrand theorem \cite{Ber} states that up to an
arbitrary factor there are only two central potentials in
Euclidean space that make all bounded trajectories of a one-body
problem closed. In spaces $\mathbf{S}^{2},\,\mathbf{H}^{2}$ also
there are only two potentials $V_{c}$ and $V_{o}$ with this
property. It was proved by H.~Liebman in 1903 \cite{Lib03}, see
also \cite{Lib}.

One can consider the classical mechanics in spaces of constant
curvature as a predecessor of special and general relativity.
After the rise of these theories the above papers of E.~Schering,
W.~Killing, H.~Liebmann were almost completely forgotten. Note
that the description of a particle motion in central potentials in
spaces $\mathbf{S}^{3}$ and $\mathbf{H}^{3}$ was shorten in the
second and the third editions of the H.~Liebman book \cite{Lib}
w.r.t.\ the first edition in favor of special relativity.

Similar models attracted attention later from the point of view of
quantum mechanics and the theory of integrable dynamical systems.
This leads to the rediscovery of results described above in many
papers. Note however that almost forgotten results of E.~Schering,
W.~Killing and H.~Liebmann were described in the survey
\cite{DomZitt}.

Quantum mechanical spectral problem on the sphere $\mathbf{S}^{3}$
for potential $V_{c}$ (Coulomb problem) was solved by
E.~Schr\"odinger in 1940 by the factorization (ladder) method,
invented by himself \cite{Schr}. A.F.~Stevenson in 1941 solved the
same problem using more traditional analysis of the hypergeometric
differential equation \cite{Ste} (see also L.~Infeld result in
1941 \cite{Inf1}). L.~Infeld and A.~Schild in 1945 solved a
similar problem in the space $\mathbf{H}^{3}$ \cite{Inf2} (see
also \cite{Inf3}).

The connection of the Runge-Lenz operator for the quantum Kepler
problem in $\mathbf{S}^{3}$ with the Schr\"{o}dinger ladder method
was discussed by A.O.~Barut and R.~Wilson in \cite{BarWil}. In
\cite{Bar} A.O.~Barut, A.~Inomata and G.~Junker solved the Kepler
problem in $\mathbf{S}^{3}$ and $\mathbf{H}^{3}$ using the
functional integration.

In papers \cite{Otch1}, \cite{Otch2} V.S.~Otchik considered the
one particle quantum two center Coulomb problem in
$\mathbf{S}^{3}$ and found a coordinate system admitting the
variable separation. The corresponding ordinary differential
equations are Heun's ones.

In \cite{Higgs} -- \cite{Gr2} there was developed an algebraic
approach to one particle problems for potentials $V_{c}$ and
$V_{o}$ in spaces $\mathbf{S}^{n},\mathbf{H}^{n}$.

Transformations between the Coulomb-Kepler and oscillator problems
existing in the Euclidean space were generalized for the sphere in
\cite{KMP}.

In \cite{BKO2} A.A.~Bogush, Yu.A.~Kurochkin and V.S.~Otchik
considered the Coulomb scattering in the space $\mathbf{H}^{3}$.

The two-body problem with a central interaction in constant
curvature spaces $\mathbf{S}^{n}$ and $\mathbf{H}^{n}$
considerably differs from its Euclidean analog. The variable
separation for the latter problem is trivial, while for the former
one no central potentials are known that admit a variable
separation.

The classical two-body problem with a central interaction in
constant curvature spaces was considered for the first time in
\cite{Shch98}. Its Hamiltonian reduction to the system with two
degrees of freedom was carried out by explicit coordinate
calculations. For some potentials there was proved the solvability
of the reduced problem for an infinite period of time.

In \cite{Shch99} there was studied the self-adjointness of the
quantum two-body Hamiltonian in spaces $\mathbf{S}^{2}$ and
$\mathbf{H}^{2}$ and were found in explicit form some its infinite
energy level series for the sphere $\mathbf{S}^{2}$, corresponding
to some central potentials.

Simply connected constant curvature spheres ${\bf S}^{n}$ and
hyperbolic spaces $\mathbf{H}^{n}$ are representatives of the
class of two-point homogeneous Riemannian spaces (TPHRS). Such
spaces are characterized by the property that any pair of points
can be transformed by means of an appropriate isometry to any
other pair of points with the same distance between them.
Equivalently, these spaces are characterized by the property that
the natural action of the isometry group on the unit sphere bundle
over them is transitive. The classification of TPHRS can be found
in \cite{Wolf}.

For a smooth manifold $M$ endowed with a left action of a Lie
group $G$ denote by $\Diff(M)\equiv\Diff_{G}(M)$ the algebra of
$G$-invariant differential operators on $M$ with smooth
coefficients. For a Riemannian manifold $M$ let $M_{S}$ be the
unit sphere bundle over $M$. Let $Q$ be an arbitrary TPHRS,
endowed with the action of the identity component of the isometry
group for $Q$.

In \cite{Shchep023} there was found a polynomial expression for
the quantum two-body Hamiltonian $H$ on $Q$ through a radial
differential operator and generators of the algebra
$\Diff(Q_{S})$. Coefficients of this polynomial depend only on the
distance between particles.

Algebras $\Diff(Q_{S})$ are noncommutative. A full set of their
generators and corresponding relations\footnote{One relation for
the quaternion projective space and its hyperbolic analog was
calculated only in leading terms.} was found in \cite{Shch022}.

Let $\mathfrak{A}$ be a set of $\Diff(Q_{S})$ generators presented
in the expression for the Hamiltonian $H$. An every common
eigenfunction of operators from $\mathfrak{A}$ generates a
separate spectral ordinary differential equation for the two-body
quantum mechanical problem on TPHRS. The search for such common
eigenfunction is not an easy problem. In low dimensions for
$Q={\bf S}^{2},\,Q={\bf S}^{3}$ this problem was solved in
\cite{Shch99} and \cite{ShchStep} using an explicit description of
$\SO(3)$ and $\SO(4)$ irreducible representations. The present
paper deals with this problem for the general spherical case
$Q={\bf S}^{n}$. The progress is reached using the results in
representations theory of the algebras
$\mathfrak{so}(n,\mathbb{C})$ in \cite{Mol1} and \cite{Mol2}

The paper is organized as follows. Sections
\ref{InvDiffOperators}--\ref{TwoBodyHamiltonian} are of a
preparatory character. Section \ref{InvDiffOperators} and
\ref{RegularRepresentations} contain basic facts on invariant
differential operators on homogeneous spaces and regular
representations of compact Lie groups respectively. In section
\ref{TwoBodyHamiltonian} there is a description of the quantum
two-body Hamiltonian on the sphere ${\bf S}^{n}$ through a radial
differential operator and generators $D_{i},\,i=0,1,2,3$ of the
algebra $\Diff\left({\bf S}^{n}_{S}\right)$.

Sections \ref{OperAction} and \ref{SpectralEquationTBP} form a
main part of the paper. In section \ref{OperAction} we calculate
actions of operators $D_{i},\,i=0,1,2,3$ in a corresponding
functional space and find all common eigenvectors $\psi_{D}$ for
operators $D_{0}^{2},D_{1},D_{2}$ and optionally $D_{3}$. Using
these eigenvectors we derive in section \ref{SpectralEquationTBP}
a separate ordinary differential equation of the second order for
a radial part of a two-body eigenfunctions. For Coulomb and
oscillator potentials this differential equation is Fuchsian and
we consider its reducibility to the hypergeometric one using the
rational change of an independent variable. This reduction is
possible for some eigenvectors $\psi_{D}$ that leads to an
explicit form of some infinite energy level series for the
two-body problem with Coulomb and oscillator potentials.

A necessary information concerning complex orthogonal Lie
algebras, self-adjoint Schr\"{o}\-din\-ger operators on Riemannian
spaces and Fuchsian differential equations is collected in
appendices \ref{Appendix A}--\ref{AppendixD}.

{\bf ACKNOWLEDGEMENT}. The author is grateful to A.I.~Molev for
pointing out his papers \cite{Mol1} and \cite{Mol2}.

\section{Invariant differential operators on homogeneous \\ spaces}
\label{InvDiffOperators}\markright{\ref{InvDiffOperators}
Invariant differential operators}

Here we shall briefly describe the construction of invariant
differential operators on homogeneous spaces \cite{Hel}.

Let $G$ be a Lie group of dimension $N$ and $K$ be its subgroup of
dimension $N-\ell$. Denote the corresponding Lie algebras as
$\mathfrak{g}$ and $\mathfrak{k}$. Suppose that the algebra
$\mathfrak{g}$ admits the reductive expansion
\begin{equation}\label{expansion}
\mathfrak{g}=\mathfrak{p}\oplus\mathfrak{k},
\end{equation}
for a subspace $\mathfrak{p}\subset\mathfrak{g}$, i.e.\
$\Ad_{K}\mathfrak{p}\subset\mathfrak{p}$. For a compact Lie group
$G$ such subspace $\mathfrak{p}$ can always be constructed using
the invariant integration on $G$. Let $(e_{j})_{j=1}^{N}$ be a
base in $\mathfrak{g}$ such that $(e_{j})_{j=1}^{\ell}$ is a base
in $\mathfrak{p}$.

Let $S(\mathfrak{p})$ be a symmetric algebra for the linear space
$\mathfrak{p}$. The $\Ad_{K}$-action on $\mathfrak{p}$ is
naturally extended to the $\Ad_{K}$-action on $S(\mathfrak{p})$.
The main result of the general theory \cite{Hel} is that
$G$-invariant differential operators on $G/K$ are in one to one
correspondence with the set $S(\mathfrak{p})^{K}$ of all
$\Ad_{K}$-invariant elements in $S(\mathfrak{p})$.

Let $\imath:\;\mathfrak{p}\to S(\mathfrak{p})$ be an inclusion,
$U(\mathfrak{g})$ be the universal enveloping algebra for
$\mathfrak{g}$ and $\lambda:\,S(\mathfrak{p})\rightarrow
U(\mathfrak{g})$ be a linear symmetrization map, defined on
monomials by the formula
$$
\lambda(e_{i_{1}}^{*}\cdot\ldots\cdot
e_{i_{k}}^{*})=\frac1{k!}\sum_{\sigma\in\mathfrak{S}_{k}}e_{i_{\sigma(1)}}\cdot\ldots\cdot
e_{i_{\sigma(k)}},\,1\leqslant i_{j}\leqslant\ell,\,1\leqslant
j\leqslant k,
$$
where $e_{i_{j}}^{*}:=\imath(e_{i_{j}})$ and $\mathfrak{S}_{k}$ is
the full permutations group of $k$ elements. Evidently
$$\lambda:\,S(\mathfrak{p})^{K}\rightarrow U(\mathfrak{g})^{K},$$
where $U(\mathfrak{g})^{K}$ is the set of all $\Ad_{K}$-invariant
elements in $U(\mathfrak{g})^{K}$.

Let $P(e_{1},\ldots,e_{N})$ be a polynomial depending on
noncommutative elements. Denote by $\tilde e_{i}$ the left
invariant vector field on $G$, corresponding to the element
$e_{i}\in\mathfrak{g}\cong T_{e}G:$
$$
\left.\tilde
e_{i}\right|_{g}=\left.\frac{d}{dt}\right|_{t=0}g\exp(te_{i}),\,g\in
G.
$$
Then $D_{P}:=P(\tilde e_{1},\ldots,\tilde e_{N})$ is a left
invariant differential operator on $G$.

Functions on the homogeneous space $G/K$ are in one to one
correspondence with functions on the group $G$ that are invariant
w.r.t.\ right $K$-shifts. For $P(e_{1},\ldots,e_{N})\in
U(\mathfrak{g})^{K}$ the differential operator $D_{P}$, acting on
such functions, can be considered as a $G$-invariant differential
operator on the space $G/K$ and every such operator can be
uniquely represented in the form
$$
\left(\lambda(P_{0})\right)(\tilde e_{1},\ldots,\tilde e_{\ell})),
$$
for some $P_{0}\in S(\mathfrak{p})^{K}$.

\section{Regular representations of compact Lie groups}\label{RepComGroups}
\label{RegularRepresentations}\markright{\ref{RegularRepresentations}
Regular representations of compact Lie groups}

Let $G$ be a compact connected Lie group and $\mu$ be a
biinvariant positive measure on $G$, unique up to an arbitrary
factor \cite{Ki1}. Let $\mathcal{L}^{2}(G,\mu)$ be a Hilbert space
of measurable complex valued functions on $G$, square integrable
w.r.t.\ the measure $\mu$. Define two unitary left representations
of $G$ in the space $\mathcal{L}^{2}(G,\mu)$. The {\it left
regular representation\label{LeftRegRep}} $T^{l}$ acts by the left
shifts
$$\left(T^{l}_{q}f\right)(g)=f(q^{-1}g),\,q,g\in G,f\in\mathcal{L}^{2}(G,\mu)$$
and the {\it right regular representation\label{RightRegRep}}
$T^{r}$ acts by the right shifts
$$\left(T^{r}_{q}f\right)(g)=f(gq),\,q,g\in G,f\in\mathcal{L}^{2}(G,\mu).$$
Evidently these representations are equivalent with the
intertwining operator $f(g)\rightarrow f(g^{-1})$. It is well
known that these representations expand into direct sums of finite
dimensional unitary irreducible representations (irreps). Each of
these irreps is contained in $T^{l}$ or $T^{r}$ with a
multiplicity equal to its dimension and an every linear
irreducible representation of $G$ is equivalent to an irreps from
this sum \cite{Ba}, \cite{Vi}.

Let $T_{\ell}$ be a full system of unitary irreps for $G$ in
spaces $U_{\ell},\,\ell=1,2,\ldots$. Choose in every $U_{\ell}$ an
orthonormal base
$(e_{\ell,k})_{k=1}^{d_{\ell}},\,d_{\ell}:=\dim_{\mathbb{C}}U_{\ell}$.
Define matrix elements  $t_{\ell,k}^{i}$ of operators $T_{q}^{r}$
by the equation $T_{q}^{r}e_{\ell,k}=:t_{\ell,k}^{i}(q)e_{\ell,i}$
or equivalently by $t_{\ell,k}^{i}(q):=\langle
e_{\ell,i},T_{q}^{r}e_{\ell,k}\rangle_{U_{\ell}},\,q\in G$. Since
$$
t_{\ell,k}^{i}(gq)e_{\ell,i}=T_{g}^{r}T_{q}^{r}e_{\ell,k}=t_{\ell,i}^{j}(g)t_{\ell,k}^{i}(q)
e_{\ell,j},\; g,q\in G
$$
one has
\begin{equation}
\label{MatrixElementAction}
t_{\ell,k}^{i}(gq)=t_{\ell,j}^{i}(g)t_{\ell,k}^{j}(q).
\end{equation}
Therefore the subspace
$\mathcal{R}_{\ell,i}\subset\mathcal{L}^{2}(G,\mu)$, spanned by
functions $(t_{\ell,j}^{i}(g))_{j=1}^{d_{\ell}}$, is invariant
under operators $T^{r}_{q}$ and the representation
$\left.T^{r}\right|_{\mathcal{R}_{\ell,i}}$ is equivalent to
$T_{\ell}$. On the other hand, the formula
(\ref{MatrixElementAction}) implies that the subspace
$\mathcal{L}_{\ell,j}\subset\mathcal{L}^{2}(G,\mu)$, spanned by
functions $(t_{\ell,j}^{i}(g))_{i=1}^{d_{\ell}}$, is invariant
under operators $T^{l}_{q}$ and the representation
$\left.T^{l}\right|_{\mathcal{L}_{\ell,j}}$ is again equivalent to
$T_{\ell}$. The functions
$(t_{\ell,j}^{i}(g))_{i,j=1}^{d_{\ell}},\,\ell=1,2,\ldots$ form an
orthogonal base in the space $\mathcal{L}^{2}(G,\mu)$ \cite{Ki1},
\cite{Ba}, \cite{Vi} and
$$
\|t_{\ell,j}^{i}\|^{2}_{\mathcal{L}^{2}(G,\mu)}=\frac{\mu(G)}{d_{\ell}}.
$$
Thus the space
$$\mathcal{T}_{\ell}:=\bigoplus_{i=1}^{d_{\ell}}\mathcal{R}_{\ell,i}=
\bigoplus_{j=1}^{d_{\ell}}\mathcal{L}_{\ell,j}
$$
is invariant under representations $T^{r}$ and $T^{l}$. The
representation $T^{r}$ intermixes spaces $\mathcal{L}_{\ell,j}$ of
representations $T^{l}$ and vise versa the representation $T^{l}$
intermixes spaces $\mathcal{R}_{\ell,i}$ of representations
$T^{r}$. The space $\mathcal{L}^{2}(G,\mu)$ of representations
$T^{r}$ and $T^{l}$ expands into irreps as follows
$$
\mathcal{L}^{2}(G,\mu)=\bigoplus_{\ell}\mathcal{T}_{\ell}=
\bigoplus_{\ell}\bigoplus_{i=1}^{d_{\ell}}\mathcal{R}_{\ell,i}
=\bigoplus_{\ell}\bigoplus_{j=1}^{d_{\ell}}\mathcal{L}_{\ell,j}.
$$
For a Lie subgroup $K$ of the group $G$ the subspace
$\mathcal{L}^{2}\left(G,K,\mu\right)\subset\mathcal{L}^{2}(G,\mu)$,
consisting of functions invariant w.r.t.\ all right $K$-shifts on
$G$, is invariant w.r.t left $G$-shifts. Therefore there are only
two possibilities:
\begin{equation*}
\mathcal{L}_{\ell,j}\cap\mathcal{L}^{2}\left(G,K,\mu\right)=\mathcal{L}_{\ell,j}\;\;
\text{and}\;\;\mathcal{L}_{\ell,j}\cap\mathcal{L}^{2}\left(G,K,\mu\right)=0.
\end{equation*}

The consideration above implies the following proposition.
\begin{proposit}\label{DimK0}
Let
\begin{equation*}
\widetilde{\mathcal{T}}_{\ell}:=\mathcal{T}_{\ell}\cap\mathcal{L}^{2}\left(G,K,\mu\right),\,
\widetilde{\mathcal{R}}_{\ell,i}:=\mathcal{R}_{\ell,i}\cap\mathcal{L}^{2}
\left(G,K,\mu\right),\,\tilde
d_{\ell}:=\dim_{\mathbb{C}}\widetilde{\mathcal{R}}_{\ell,i}.
\end{equation*}
Evidently, the value $\tilde d_{\ell}$ does not depend on
$i=1,\ldots,d_{\ell}$. The representation
$\left.T^{l}\right|_{\widetilde{\mathcal{T}}_{\ell}}$ is expanded
into the direct sum of equivalent irreps in spaces
$\mathcal{L}_{\ell,k}^{K},\,k=1,\ldots,\tilde d_{\ell}$, which are
among of $\mathcal{L}_{\ell,j}$. On the other hand
$$
\widetilde{\mathcal{T}}_{\ell}=\bigoplus_{i=1}^{d_{\ell}}\widetilde{\mathcal{R}}_{\ell,i},
$$
where the spaces
$\widetilde{\mathcal{R}}_{\ell,i},\,i=1,\ldots,d_{\ell}$ are
isomorphic to each other.
\end{proposit}

\section{Two-body Hamiltonian on the sphere ${\bf S}^{n}$}
\label{TwoBodyHamiltonian}\markright{\ref{TwoBodyHamiltonian}
Two-body Hamiltonian on the sphere ${\bf S}^{n}$}

Let ${\bf S}^{n},\,n\geqslant2$ be the $n$-dimensional sphere,
endowed with the standard metric $g$ of a constant sectional
curvature $R^{-2},\,R>0$ and
$$
\bigtriangleup=\frac1{\sqrt{\gamma}}
\pd1{x^{i}}\left(\sqrt{\gamma}g^{ij}\pd1{x^{j}}\right),
$$
be the corresponding Laplace-Beltrami operator, expressed through
local coordinates, where $\gamma:=\det\|g_{ij}\|$. We start from
the description of the two-body quantum Hamiltonian on ${\bf
S}^{n}$ found in \cite{Shchep023} and \cite{Sh001}.

The configurations space for the two-body system on ${\bf S}^{n}$
is
\begin{equation}\label{ConfSpace}
{\bf S}^{n}\times{\bf S}^{n}.
\end{equation}
The Hamiltonian for this system is
\begin{equation}\label{Ham}
H_{V}=H_{0}+V\equiv-\frac1{2m_{1}}\bigtriangleup_{1}-\frac1{2m_{2}}\bigtriangleup_{2}+V(\rho),
\end{equation}
where $\bigtriangleup_{i},\,i=1,2$ is the Laplace-Beltrami
operator on the ith factor of (\ref{ConfSpace}) and $\rho$ be the
distance between particles. It should be defined on some subspace
$\Dom(H)$ dense in $\mathcal{L}^{2}\left({\bf S}^{n}\times{\bf
S}^{n},\chi\times\chi\right)$ to be a self-adjoint operator, where
$\chi$ is the measure on ${\bf S}^{n}$ induced by the metric. In
local coordinates $\chi$ has the form:
$\chi=\sqrt{\gamma}dx_{1}\wedge\ldots\wedge dx_{n}$. Note that the
free Hamiltonian $H_{0}$ is the Laplace-Beltrami operator for the
metric
\begin{equation}\label{g2metric}
g_{2}:=m_{1}\tilde\pi^{*}_{1}g + m_{2}\tilde\pi^{*}_{2}g
\end{equation}
on (\ref{ConfSpace}), multiplied by $-1/2$, where
$\tilde\pi^{*}_{i}g$ is the pullback of the metric $g$ with
respect to the projection on the $i$-th factor.

Let $G\cong\SO(n+1)$ be the identity component of the isometry
group for the sphere ${\bf S}^{n}$. One can consider $\SO(n+1)$ in
the standard way as
$$
\SO(n+1)=\left(\left.A\in\GL(n+1,\mathbb{R})\right|\;AA^{T}=E,\,\det
A=1\right),
$$
where $E$ is the matrix unit. The configuration space
(\ref{ConfSpace}) is endowed with the diagonal $G$-action and the
differential operator (\ref{Ham}) is $G$-invariant.

Let $K\cong\SO(n-1)$ be a subgroup in $ \SO(n+1)$ with elements of
the form
$$
\left(\begin{array}{cc} E_{2} & 0 \\ 0 & A
\end{array}\right),\;E_{2}=\left(\begin{array}{cc} 1 & 0 \\ 0 & 1
\end{array}\right),\,A\in\SO(n-1).
$$

Up to a manifold of dimension $n$, consisting of antipodal points,
the configuration space (\ref{ConfSpace}) can be represented as
the direct product
\begin{equation}
I\times\left(G/K\right),
\end{equation}
where $I=(0,\pi R)$ and the factor space $G/K$ is $G$-homogeneous
w.r.t.\ left shifts \cite{Shchep023}. The space $G/K$ is
isomorphic to the unit sphere bundle over ${\bf S}^{n}$
\cite{Shch022}.

The Lie algebra $\mathfrak{g}\cong\mathfrak{so}(n+1)$,
corresponding to the group $G$, consists of skew-symmetric
matrices. Let $E_{kj}$ be the matrix of the size
$(n+1)\times(n+1)$ with the unique nonzero element equals $1$,
locating at the intersection of the $k$-th row and the $j$-th
column. Choose the base for the algebra $\mathfrak{g}$ as:
\begin{align*}
\Psi_{kj}=E_{kj}-E_{jk},\,1\leqslant k<j\leqslant n+1.
\end{align*}
The algebra $\mathfrak{g}$ admits the reductive expansion
(\ref{expansion}), where the subspace $\mathfrak{p}$ is spanned by
elements
$$
\Psi_{1k},\,2\leqslant k\leqslant n+1,\;\Psi_{2k},\,3\leqslant
k\leqslant n+1.
$$

In the general case $n\geqslant4$ generators of the commutative
algebra $S(\mathfrak{p})^{K}$ can be chosen \cite{Shch022} as:
\begin{equation*}
-\Psi_{12}^{*},\,\sum\limits_{k=3}^{n+1}\left(\Psi_{1k}^{*}\right)^{2},\,
\sum\limits_{k=3}^{n+1}\left(\Psi_{2k}^{*}\right)^{2},\,
-\sum\limits_{k=3}^{n+1}\Psi_{1k}^{*}\Psi_{2k}^{*}.
\end{equation*}

In the case $n=3$ there is the additional generator
$$
\square^{*}=\Psi_{13}^{*}\Psi_{24}^{*}-\Psi_{14}^{*}\Psi_{23}^{*}.
$$
In the case $n=2$ the group $K$ is trivial and generators of
$S(\mathfrak{p})^{K}=S(\mathfrak{p})$ are simply
$$
\Psi_{12}^{*},\,\Psi_{13}^{*},\,\Psi_{23}^{*}.
$$

In all cases we shall consider elements
\begin{equation}\label{generators}
D_{0}=-\Psi_{12},\,D_{1}=\sum\limits_{k=3}^{n+1}\Psi_{1k}^{2},\,
D_{2}=\sum\limits_{k=3}^{n+1}\Psi_{2k}^{2},\,
D_{3}=-\frac12\sum\limits_{k=3}^{n+1}\left\{\Psi_{1k},\Psi_{2k}
\right\},
\end{equation}
from $U(\mathfrak{g})$ as invariant differential operators on the
space $G/K$, where $\{\cdot,\cdot\}$ means an anticommutator. The
commutative relations for differential operators
(\ref{generators}) are (see \cite{Shch022})
\begin{align}\label{CommRel}
&[D_{0},D_{1}]=-2D_{3},\,[D_{0},D_{2}]=2D_{3},\,[D_{0},D_{3}]=D_{1}-D_{2},\,
[D_{1},D_{2}]=-2\{D_{0},D_{3}\},\\ &[D_{1},D_{3}]=-\{D_{0},D_{1}\}
+\frac{(n-1)(n-3)}2D_{0},\,[D_{2},D_{3}]=\{D_{0},D_{2}\}-\frac{(n-1)(n-3)}2D_{0}.\notag
\end{align} For $n=3$ the additional operator
$$
\square:=\frac12\left(\{\Psi_{13},\Psi_{24}\}-
\{\Psi_{14},\Psi_{23}\}\right)
$$
lies in the centre of
the algebra $\Diff_{G}(G/K)$.

Define a new coordinate $r$ on the interval $I$ by the equation
$$r=\tan\left(\frac{\rho}{2R}\right),\,r\in\mathbb{R}_{+}:=(0,\infty).$$

Results from \cite{Shchep023} and \cite{Shch022} imply the
following theorem.
\begin{theore}\label{TwoBodyHamComTh}
The quantum two-body Hamiltonian on the sphere ${\bf S}^{n}$ can
be considered as the differential operator
\begin{align}\label{TBHamR}
H&=-\frac{(1+r^2)^{n}}{8mR^2r^{n-1}}\pd1{r}\circ\left(
\frac{r^{n-1}}{(1+r^2)^{n-2}}\pd1{r}\right)-\frac{m_{1}\alpha^{2}+m_{2}\beta^{2}}{2m_{1}m_{2}R^{2}}D_{0}^{2}
\\ &+
\frac{(m_{1}\alpha-m_{2}\beta)(1+r^{2})^{n}}{4m_{1}m_{2}R^{2}
r^{n-1}}
\left\{\pd1{r},\frac{r^{n-1}D_{0}}{(1+r^{2})^{n-1}}\right\}-
\frac12\left(C D_{1}+A D_{2}+2B D_{3}\right)+V(r),\notag
\end{align}
on the space $\mathbb{R}_{+}\times G$, where
\begin{equation}\label{RedMass}
m:=\frac{m_{1}m_{2}}{m_{1}+m_{2}},
\end{equation} a parameter
$\alpha\in (0,1)$ is arbitrary, $\beta:=1-\alpha$, and
\begin{align*}
A&=\frac{(1+r^{2})^{2}}{4m_{1}m_{2}R^{2}r^{2}}\left(m_1\cos^{2}(2\alpha\arctan
r)+ m_2\cos^{2}(2\beta\arctan r)\right),\\
B&=\frac{(1+r^{2})^{2}}{8m_{1}m_{2}R^{2}r^{2}}\left(m_1\sin(4\alpha\arctan
r)- m_2\sin(4\beta\arctan r)\right),\\
C&=\frac{(1+r^{2})^{2}}{4m_{1}m_{2}R^{2}r^{2}}\left(m_1\sin^{2}(2\alpha\arctan
r)+ m_2\sin^{2}(2\beta\arctan r)\right).
\end{align*}
The domain for operator (\ref{TBHamR}) is dense in the space
$\mathcal{L}^{2}\left(\mathbb{R}_{+}\times G,K,\eta\right)$,
consisting of all complex valued square integrable $K$-invariant
functions on $\mathbb{R}_{+}\times G$, with respect to right
$K$-shifts and the measure
$$
d\eta=\frac{r^{n-1}dr}{(1+r^{2})^{n}}\otimes d\mu\equiv
d\nu\otimes d\mu,
$$
where $\mu$ is a biinvariant measure on $G$, unique up to a
constant factor.
\end{theore}
In the following we choose the parameter $\alpha$ in such a way
that $m_{1}\alpha-m_{2}\beta=0$, i.e.\
$$
\alpha=\frac{m_{2}}{m_{1}+m_{2}},\,\beta=\frac{m_{1}}{m_{1}+m_{2}}.
$$
For such choice operator (\ref{TBHamR}) becomes
\begin{align}
\begin{split}
H&=-\frac{(1+r^2)^{n}}{8mR^2r^{n-1}}\pd1{r}\circ\left(
\frac{r^{n-1}}{(1+r^2)^{n-2}}\pd1{r}\right)-\frac1{2(m_{1}+m_{2})R^{2}}D_{0}^{2}
\\ &-
\frac12\left(C D_{1}+A D_{2}+2B D_{3}\right)+V(r).
\end{split}\label{TBHamR1}
\end{align}

It is obvious that
\begin{equation}\label{SpaceFactorization}
\mathcal{L}^{2}\left(\mathbb{R}_{+}\times
G,K,\eta\right)=\mathcal{L}^{2}\left(\mathbb{R}_{+},\nu\right)\otimes
\mathcal{L}^{2}\left(G,K,\mu\right).
\end{equation}
Operators $D_{0}^{2},D_{1},D_{2},D_{3}$ act on the second factor
in (\ref{SpaceFactorization}). This action will be studied in the
following section.

Note that $B\equiv0$ for $m_{1}=m_{2}$. Let $\psi_{D}\in
\mathcal{L}^{2}\left(G,K,\mu\right)$ be a common eigenfunctions
for operators $D_{0}^{2},D_{1},D_{2}$ if $m_{1}=m_{2}$ and also
for $D_{3}$ if $m_{1}\ne m_{2}$. Then the following stationary
Schr\"{o}dinger equation
\begin{equation}\label{StatSchrEq}
H\left(f(r)\psi_{D}\right)=Ef(r)\psi_{D}
\end{equation}
is equivalent to a spectral problem for an ordinary differential
equation for a function $f(r)$ and an energy level $E$ (in other
words to a one-dimensional stationary Shr\"{o}dinger
equation).\footnote{Note that such eigenfunctions are very special
elements of the space $\mathcal{L}^{2}\left(G,K,\mu\right)$ and
they do not span it.}

\begin{proposit}\label{aprioriInf}
Let $\psi_{D}$ be a common eigenfunction for operators
$D_{0}^{2},D_{1},D_{2},D_{3}$ with eigenvalues
$\delta_{0},\delta_{1},\delta_{2}$ and $\delta_{3}$ respectively.
Then
\begin{enumerate}
\item $\delta_{1}=\delta_{2}$ and $\delta_{3}=0$;
\item $D_{0}\psi_{D}$ is an eigenfunction for operators
$D_{0}^{2},D_{1},D_{2},D_{3}$ with the same eigenvalues
$\delta_{0},\delta_{1},\delta_{2}$ and $\delta_{3}$ respectively;
\item if $D_{0}\psi_{D}\not\sim\psi_{D}$, then
$D_{0}\psi_{D}\pm\sqrt{\delta_{0}}\psi_{D}$ are eigenfunctions for
operators $D_{0},D_{1},D_{2},D_{3}$;
\item if $D_{0}\psi_{D}\sim\psi_{D}$, then either $D_{0}\psi_{D}=0$
or $\delta_{1}=\delta_{2}=(n-1)(n-3)/4$.
\end{enumerate}
\end{proposit}
\begin{proof}
Relations $[D_{0},D_{3}]=D_{1}-D_{2}$ and
$[D_{1},D_{2}]=-2\{D_{0},D_{3}\}$ imply
\begin{align}
[D_{0},D_{3}]\psi_{D}&=\delta_{3}D_{0}\psi_{D}-D_{3}D_{0}\psi_{D}=(D_{1}-D_{2})\psi_{D}=
(\delta_{1}-\delta_{2})\psi_{D},\notag\\
\label{PsiD1}\delta_{3}D_{0}\psi_{D}&+D_{3}D_{0}\psi_{D}=\{D_{0},D_{3}\}\psi_{D}=
-\frac12[D_{1},D_{2}]\psi_{D}=0.
\end{align}
The last two equations lead to
\begin{equation}\label{PsiD2}
2\delta_{3}D_{0}\psi_{D}=(\delta_{1}-\delta_{2})\psi_{D}.
\end{equation}
If $\delta_{3}\neq0$, then the last equation implies
$D_{0}\psi_{D}\sim\psi_{D}$ and the relation
$[D_{0},D_{1}]=-2D_{3}$ gives
$\delta_{3}\psi_{D}=D_{3}\psi_{D}=-\frac12[D_{0},D_{1}]\psi_{D}=0$.
Thus $\delta_{3}=0$ and equation (\ref{PsiD2}) implies
$\delta_{1}=\delta_{2}$ that proves the first claim of the
proposition.

Now from equation (\ref{PsiD1}) one gets $D_{3}D_{0}\psi_{D}=0$
and the first two relations (\ref{CommRel}) imply
$D_{1}D_{0}\psi_{D}=D_{2}D_{0}\psi_{D}=\delta_{1}D_{0}\psi_{D}$.
The relation $D_{0}^{2}D_{0}\psi_{D}=\delta_{0}D_{0}\psi_{D}$ is
evident, which completes the proof of the second claim.

The relation $D_{0}^{2}\psi_{D}=\delta_{0}\psi_{D}$ is equivalent
to
$\left(D_{0}+\sqrt{\delta_{0}}\id\right)\left(D_{0}-\sqrt{\delta_{0}}\id\right)\psi_{D}=0$.
Now if $D_{0}\psi_{D}\neq\sqrt{\delta_{0}}\psi_{D}$, then
$\psi_{D}^{-}:=\left(D_{0}-\sqrt{\delta_{0}}\id\right)\psi_{D}$ is
an eigenfunction for the operator $D_{0}$. The function
$\psi_{D}^{-}$ is also an eigenfunction for operators
$D_{1},D_{2},D_{3}$ due to the second claim. The consideration for
the function
$\psi_{D}^{+}:=\left(D_{0}+\sqrt{\delta_{0}}\id\right)\psi_{D}$ is
completely similar. Thus the third claim is proved.

Assume now $D_{0}\psi_{D}=\delta_{0}'\psi_{D}$. Then the last
relation from (\ref{CommRel}) gives
$$
2\delta_{0}'\delta_{2}\psi_{D}=\frac12(n-1)(n-3)\delta_{0}'\psi_{D}.
$$
It means either $\delta_{0}'=0$ or
$\delta_{1}=\delta_{2}=(n-1)(n-3)/4$ that proves the last claim.
\end{proof}

\section{Action of operators $D_{0},D_{1},D_{2},D_{3}$ in the
space \\
$\mathcal{L}^{2}\left(G,K,\mu\right)$}\label{OperAction}\markright{\ref{OperAction}
Action of operators $D_{0}^{2},D_{1},D_{2},D_{3}$.}

Here we use notations of section \ref{RepComGroups} for
$G=\SO(n+1)$ and $K=\SO(n-1)$. Below we mean by the
complexification $\mathfrak{g}^{\mathbb{C}}$ of the Lie algebra
$\mathfrak{g}$ the following set
\begin{equation}\label{StandardForm}
\mathfrak{so}(n+1,\mathbb{C})=\left(\left.A\in\mathfrak{gl}(n+1,\mathbb{C})\right|\;
A+A^{T}=E\right).
\end{equation}

Operators $D_{i}$ are polynomial w.r.t.\ infinitesimal generators
of right $G$-shifts. Therefore they conserve the spaces
$\widetilde{\mathcal{T}}_{\ell}$ and generally intermix its direct
summands $\mathcal{L}_{\ell,k}^{K},\,k=1,\ldots,\tilde d_{\ell}$
with constant $\ell$ and different $k$. On the other hand they act
in spaces $\widetilde{R}_{\ell,i}$ and their action is the same
for constant $\ell$ and different $i=1,\ldots,d_{\ell}$.

From now we shall treat complex spaces $R_{\ell,i}$ as a simple
left modules over $\mathfrak{g}^{\mathbb{C}}$. Their subspaces
$\widetilde{R}_{\ell,i}$ consist of elements annulled by the
subalgebra
$\mathfrak{k}^{\mathbb{C}}\cong\mathfrak{so}(n-1,\mathbb{C})\subset\mathfrak{g}^{\mathbb{C}}$.

The classification of such modules based on the notion of a
dominant weight is well-known \cite{Hamphreys}, \cite{GotoGross}
(see also appendices A and B for a brief description). In order to
apply this theory one should use a form of
$\mathfrak{so}(n+1,\mathbb{C})$, described in appendix
\ref{Appendix A} and different from  (\ref{StandardForm}).
Besides, since $\mathfrak{B}_{k}:=\mathfrak{so}(2k+1,\mathbb{C})$
and $\mathfrak{D}_{k}:=\mathfrak{so}(2k,\mathbb{C})$ are different
series of simple complex Lie algebras, we shall consider cases of
odd and even $n$ separately.

\subsection{The case $n=2k$}\label{n=2k}

In this section we shall use notations from appendix
\ref{BCaseAp}. In particular, by $\mathfrak{B}_{k}$ we mean the
set (\ref{BkForm}). First of all we shall construct the
isomorphism $\mathfrak{g}^{\mathbb{C}}\cong\mathfrak{B}_{k}$ in
explicit form.

Let
$$
J_{2k+1}=\begin{pmatrix} \frac1{\sqrt{2}}E_{k} & 0 &
\frac1{\sqrt{2}}S_{k} \\ 0 & 1 & 0 \\ \frac{\ii}{\sqrt{2}}S_{k} &
0 & \frac{-\ii}{\sqrt{2}}E_{k}
\end{pmatrix}\in\GL(2k+1,\mathbb{C}),
$$
where $\ii$ is the complex unit. It is easily verified that
$$
J_{2k+1}S_{2k+1}J^{T}_{2k+1}=E_{2k+1}.
$$
Therefore the equation $A^{T}S_{2k+1}+S_{2k+1}A=0$ for
$A\in\mathfrak{gl}(2k+1,\mathbb{C})$ is equivalent to the equation
$B^{T}+B=0$, where
$B:=\left(J^{T}_{2k+1}\right)^{-1}AJ^{T}_{2k+1}$. Thus the map
\begin{equation}\label{Trans1}
B\rightarrow J^{T}_{2k+1}B\left(J^{T}_{2k+1}\right)^{-1}
\end{equation}
is the isomorphism between $\mathfrak{g}^{\mathbb{C}}$ and
$\mathfrak{B}_{k}$.

Let
$$
C=\begin{pmatrix} 0 & \alpha & A_{-} & a & A_{+} \\ -\alpha & 0 &
B_{-} & b & B_{+} \\ -A_{-}^{T} & -B_{-}^{T} & & & \\ -a & -b & &
C' &
\\ -A_{+}^{T} & -B_{+}^{T} & & &
\end{pmatrix}\in\mathfrak{g},
$$
where
\begin{align*}
A_{-}&=\left(a_{-(k-1)},\ldots,a_{-1}\right),\,
A_{+}=\left(a_{1},\ldots,a_{k-1}\right),\,
B_{-}=\left(b_{-(k-1)},\ldots,b_{-1}\right),\\
B_{+}&=\left(b_{1},\ldots,b_{k-1}\right),\,a_{i},b_{i},a,b\in\mathbb{R},\,
C'\in\mathfrak{so}(2k-1).
\end{align*}
Move the second row and the second column of the matrix $C$ to the
last positions. This gives the matrix
$$
\widetilde{C}=\begin{pmatrix} 0 & A & a & A_{+} & \alpha  \\
-A_{-}^{T} & & & & -B_{-}^{T}\\ -a & & \widetilde{C}' & & -b
\\ -A_{+}^{T} & & &  & -B_{+}^{T} \\ -\alpha & B_{-} & b & B_{+} &
0\end{pmatrix}\in\mathfrak{so}(2k+1),\,\widetilde{C}'\in\mathfrak{so}(2k-1).
$$
The transformation (\ref{Trans1}) now gives for
$
\widehat{C}:=J^{T}_{2k+1}\widetilde{C}\left(J^{T}_{2k+1}\right)^{-1}$
the expression
$$
\widehat{C}=\frac12\begin{pmatrix} -2\ii\alpha & Z_{-}-\ii
Z_{+}S_{k-1} & \sqrt{2}z & Z_{-}S_{k-1}+\ii Z_{+} & 0
\\ -\overline{Z}_{-}^{T}-\ii S_{k-1}\overline{Z}_{+}^{T} & & & &
-Z_{-}^{T}-\ii S_{k-1}Z_{+}^{T} \\ -\sqrt{2}\bar z & &
\widehat{C}' & & -\sqrt{2}z \\
-S_{k-1}\overline{Z}_{-}^{T}+\ii\overline{Z}_{+}^{T} & & & &
-S_{k-1}Z_{-}^{T}+\ii Z_{+}^{T} \\ 0 & \overline{Z}_{-}-\ii
\overline{Z}_{+}S_{k-1} & \sqrt{2}\bar z &
\overline{Z}_{-}S_{k-1}+\ii\overline{Z}_{+} & 2\ii\alpha
\end{pmatrix},
$$
where $Z_{-}:=A_{-}+\ii B_{-},\,Z_{+}:=A_{+}+\ii B_{+},\,z:=a+\ii
b,\,\widehat{C}'\in\mathfrak{B}_{k-1}$. Let us identify Lie
algebras $\mathfrak{g}^{\mathbb{C}}$ and $\mathfrak{B}_{k}$
through the map $C\rightarrow\widehat{C}$. Due to the definition
of $\Psi_{ij}$ in section \ref{TwoBodyHamiltonian} one gets the
following formulas
\begin{align*}
\Psi_{12}&=\ii F_{kk},\,\Psi_{1,k+2}=
\frac1{\sqrt{2}}\left(F_{k0}-F_{0k}\right),\,\Psi_{2,k+2}=-
\frac{\ii}{\sqrt{2}}\left(F_{k0}+F_{0k}\right),\\
\Psi_{1i}&=\frac12\left(F_{kj}+F_{k,-j}+F_{-kj}+F_{-k,-j}\right),\,j=i-k-2,\,3\leqslant
i\leqslant k+1,\\ \Psi_{1i}&=\frac{\ii}2
\left(F_{kj}-F_{k,-j}+F_{-kj}-F_{-k,-j}\right),\,j=i-k-2,\,k+3\leqslant
i\leqslant 2k+1,\\
\Psi_{2i}&=\frac{\ii}2\left(F_{-kj}+F_{-k,-j}-F_{kj}-F_{k,-j}\right),\,j=i-k-2,\,3\leqslant
i\leqslant k+1,\\
\Psi_{2i}&=\frac12\left(F_{kj}-F_{k,-j}+F_{-k,-j}-F_{-kj}\right),\,j=i-k-2,\,k+3\leqslant
i\leqslant 2k+1,
\end{align*}
which imply
\begin{align}\label{DPsi}
D_{1}&=\frac12\left(F_{k0}-F_{0k}\right)^{2}+
\frac12\sum_{j=1}^{k-1}\left\{F_{-kj}+F_{kj},
F_{k,-j}+F_{-k,-j}\right\},\notag\\
D_{2}&=-\frac12\left(F_{k0}+F_{0k}\right)^{2}+
\frac12\sum_{j=1}^{k-1}\left\{F_{-kj}-F_{kj},
F_{k,-j}-F_{-k,-j}\right\},\\
D_{3}&=\frac{\ii}2\left(F_{k0}^{2}-F_{0k}^{2}\right)+
\ii\sum_{j=1}^{k-1}\left(F_{kj}F_{k,-j}-
F_{-kj}F_{-k,-j}\right),\,D_{0}=-\ii F_{kk}.\notag
\end{align}

Since the case $k=1$ does not fit the general scheme due to the
triviality of the group $K$ we assume from now $k\geqslant 2$. The
case $k=1$ will be considered below.

Let the space $\mathcal{R}_{\ell,i}$ equals
$V_{\mathfrak{B}_{k}}(\lambda)$ for a highest weight
(\ref{HigestWeightB}), where $m_{i}\in\mathbb{Z}_{+}$, and
$\widetilde{V}_{\mathfrak{B}_{k}}(\lambda)$ be a subspace of
$V_{\mathfrak{B}_{k}}(\lambda)$ annulled by the subalgebra
$\mathfrak{k}^{\mathbb{C}}\cong\mathfrak{B}_{k-1}$. An element
$v\in\widetilde{V}_{\mathfrak{B}_{k}}(\lambda),\,v\ne0$ is a
highest vector of the trivial one-dimensional
$\mathfrak{B}_{k-1}$-module. Then propositions
\ref{BtoDRestriction} and \ref{DtoBRestriction} imply the
existence of such numbers $m_{j}'\in\mathbb{Z}_{+},\,j=1,\ldots,k$
that
\begin{gather*}
m_{k}\geqslant m_{k}'\geqslant m_{k-1}\geqslant\ldots\geqslant
m_{2}'\geqslant m_{1}\geqslant m_{1}'\geqslant -m_{1},\\
m_{k}'\geqslant 0\geqslant
m_{k-1}'\geqslant0\geqslant\ldots\geqslant m_{2}'\geqslant
0\geqslant |m_{1}'|.
\end{gather*}
Thus $m_{j}'=0,\,j=1,\ldots,k-1$ and therefore
$m_{j}=0,\,j=1,\ldots,k-2$.

From now till the end of the present subsection suppose
$$
\lambda=m_{k-1}\varepsilon_{k-1}+m_{k}\varepsilon_{k},\,m_{k}\geqslant
m_{k-1}\geqslant0,\, m_{k},m_{k-1}\in\mathbb{Z}_{+}.
$$
In this case proposition \ref{BtoDRestriction} implies that an
every module $V_{\mathfrak{D}_{k}}(m_{k}'\varepsilon_{k}) \subset
V_{\mathfrak{B}_{k}}(\lambda)$ contains the unique one-dimensional
module $V_{\mathfrak{B}_{k-1}}(0)$. This fact leads to
\begin{equation}\label{DIM}
\dim\widetilde{V}_{\mathfrak{B}_{k}}(\lambda)=m_{k}-m_{k-1}+1.
\end{equation}

Thus from proposition \ref{DimK0} one gets the following expansion
\cite{Lev}:
$$
\mathcal{L}^{2}\left(\SO(2k+1),\SO(2k-1),\mu\right)=
\bigoplus_{\genfrac{}{}{0pt}{0}{m_{k}\geqslant
m_{k-1}}{m_{k},m_{k-1}\in\mathbb{Z}_{+}}}(m_{k}-m_{k-1}+1)V_{\mathfrak{B}_{k}}
\left(m_{k}\varepsilon_{k}+m_{k-1}\varepsilon_{k-1}\right),
$$
where the left hand side is considered as a restriction of the
left regular representation for the group $\SO(2k+1)$. On the
other hand the space
$$\mathcal{L}^{2}\left(\SO(2k+1),\SO(2k-1),\mu\right)$$ as a
$\Diff_{\SO(2k+1)}(\SO(2k+1)/\SO(2k-1))$-module is expanded as
\begin{align}\label{ExpanTilde}
\mathcal{L}^{2}&\left(\SO(2k+1),\SO(2k-1),\mu\right)\notag\\&=
\bigoplus_{\genfrac{}{}{0pt}{0}{m_{k}\geqslant
m_{k-1}}{m_{k},m_{k-1}\in\mathbb{Z}_{+}}}\left(\dim
V_{\mathfrak{B}_{k}}\left(m_{k}\varepsilon_{k}+m_{k-1}\varepsilon_{k-1}\right)\right)
\widetilde{V}_{\mathfrak{B}_{k}}\left(m_{k}\varepsilon_{k}+m_{k-1}\varepsilon_{k-1}\right),
\end{align} where $\dim
V_{\mathfrak{B}_{k}}\left(m_{k}\varepsilon_{k}+m_{k-1}\varepsilon_{k-1}\right)$
is given by (\ref{Weyl}).

Let
\begin{align*}
D^{+}&:=\sum_{j=1}^{k-1}F_{kj}F_{k,-j}+\frac12F_{k0}^{2},\,
D^{-}:=\sum_{j=1}^{k-1}F_{-kj}F_{-k,-j}+\frac12F_{0k}^{2},\\
\widetilde{C}&:=\left.C\right|_{\mathcal{L}^{2}\left(\SO(2k+1),\SO(2k-1),\mu\right)}
=F_{kk}^{2}+\{F_{k0},F_{0k}\}+
\sum_{j=1}^{k-1}\left(\{F_{kj},F_{jk}\}+\{F_{k,-j},F_{-jk}\}\right)
\end{align*}
be operators from $\Diff_{\SO(2k+1)}(\SO(2k+1)/\SO(2k-1))$, where
$C$ is the universal Casimir operator (\ref{Casimir}). Due to
(\ref{WeigtSubspacesB}) and (\ref{WeightMove}) the operator
$D^{+}$ "raises" weight subspaces of
$\widetilde{V}_{\mathfrak{B}_{k}}(\lambda)$ and the operator
$D^{-}$ "lowers" them.

Since $[F_{kj},F_{k,-j}]=[F_{-kj},F_{-k,-j}]=0$ one gets the
following relations
\begin{gather}
D_{1}=D^{+}+D^{-}+\frac12\left(F_{kk}^{2}-\widetilde{C}\right),\,
D_{2}=-D^{+}-D^{-}+\frac12\left(F_{kk}^{2}-\widetilde{C}\right),\,
D_{3}=\ii\left(D^{+}-D^{-}\right),\notag \\ \label{DDrelations}
D^{+}=\frac14\left(D_{1}-D_{2}\right)-\frac{\ii}2D_{3},\,
D^{-}=\frac14\left(D_{1}-D_{2}\right)+\frac{\ii}2D_{3},\,
\widetilde{C}=-D_{0}^{2}-D_{1}-D_{2}.
\end{gather}
Commutator relations (\ref{CommRel}) now give
\begin{align}\label{FDComm}
[F_{kk},D^{+}]&=2D^{+},\,[F_{kk},D^{-}]=-2D^{-},\\ \label{DDComm}
[D^{+},D^{-}]&=-\frac12F_{kk}^{3}+\frac12\widetilde{C}F_{kk}+
\frac14(2k-1)(2k-3)F_{kk}.
\end{align}
Formulas (\ref{DeltaDefB}) and (\ref{CasimirEigen}) implies
\begin{equation}\label{CasimirEigenBCase}
\left.\widetilde{C}\right|_{\widetilde{V}_{\mathfrak{B}_{k}}(\lambda)}=
\left(\left(k+m_{k}-\frac12\right)^{2}+\left(k+m_{k-1}-\frac32\right)^{2}
-\left(k-\frac12\right)^{2}-\left(k-\frac32\right)^{2}\right)\id.
\end{equation}
It follows from the paper \cite{Mol2} that
\begin{equation}\label{MainExpansion}
\widetilde{V}_{\mathfrak{B}_{k}}(\lambda)=V_{-\nu\varepsilon_{k}}\oplus
V_{-(\nu-2)\varepsilon_{k}}\oplus\ldots\oplus
V_{(\nu-2)\varepsilon_{k}}\oplus V_{\nu\varepsilon_{k}},
\end{equation}
where $\nu=m_{k}-m_{k-1}$ and all summands are one-dimensional
weight spaces w.r.t.\ the Cartan subalgebra
$\mathfrak{h}_{k}$.\footnote{In appendix \ref{appD} we shall give
a proof of expansion (\ref{MainExpansion}) independent from the
theory of Yangians used in \cite{Mol2}.} Formulas
(\ref{WeigtSubspacesB}) and (\ref{WeightMove}) imply
$$
D^{+}:\;V_{j\varepsilon_{k}}\rightarrow
V_{(j+2)\varepsilon_{k}},\;D^{-}:\;V_{j\varepsilon_{k}}\rightarrow
V_{(j-2)\varepsilon_{k}}.
$$

The action of operators $F_{kk},D^{+},D^{-}$ in the space
$\widetilde{V}_{\mathfrak{B}_{k}}(\lambda)$ was calculated in
\cite{Mol2} w.r.t.\ some base. In particular, in
$\widetilde{V}_{\mathfrak{B}_{k}}(\lambda)$ there are no
nontrivial invariant subspaces w.r.t.\ this action. We shall
obtain simpler formulas for the $D^{+}$ and $D^{-}$-action w.r.t.\
a base in $\widetilde{V}_{\mathfrak{B}_{k}}(\lambda)$ with a
normalization different from those in \cite{Mol2}.

\begin{Lem}\label{PrD-actions}
Let $L_{\nu}:=(-\nu,-\nu+2,\ldots,\nu-2,\nu)$. There is a base
$\left(\chi_{j}\right)_{j\in L_{\nu}}$ in
$\widetilde{V}_{\mathfrak{B}_{k}}(\lambda)$ such that
\begin{align}\label{DF+action}
F_{kk}\chi_{j}=j\chi_{j}&,\;
D^{+}\chi_{j}=\frac14(j-m_{k}-m_{k-1}-2k+3)(j-\nu)\chi_{j+2},\\
\label{D-action}
D^{-}\chi_{j}&=\frac14(j+m_{k}+m_{k-1}+2k-3)(j+\nu)\chi_{j-2},
\end{align}
where $\chi_{j}=0$ if $j\not\in L_{\nu}$.
\end{Lem}
\begin{proof}
Since the action of an algebra, generated by operators
$F_{kk},D^{+},D^{-}$, is irreducible in
$\widetilde{V}_{\mathfrak{B}_{k}}(\lambda)$, one can define by
induction nonzero elements $\chi_{j}\in
V_{j\varepsilon_{k}},\,j\in L_{\nu}$ such that formulas
(\ref{DF+action}) are valid. Prove by induction formula
(\ref{D-action}). For $j=-\nu$ it is evident. Suppose that
(\ref{D-action}) is valid for $j=-\nu,-\nu+2,\ldots,i$, where
$i<\nu$. Then using (\ref{CasimirEigenBCase}) one gets
\begin{align*}
&\frac14(i-m_{k}-m_{k-1}-2k+3)(i-\nu)D^{-}\chi_{i+2}=D^{-}D^{+}\chi_{i}=
\left([D^{-},D^{+}]+D^{+}D^{-}\right)\chi_{i}\\
&=\left(\frac12F_{kk}^{3}-\frac12\widetilde{C}F_{kk}-
\frac14(2k-1)(2k-3)F_{kk}\right)\chi_{i}\\
&+\frac14(i+m_{k}+m_{k-1}+2k-3)(i+\nu)D^{+}\chi_{i-2}\\&=\frac12\left(i^{3}-i\left(m_{k}^{2}
+m_{k-1}^{2}+(2k-1)m_{k}+(2k-3)m_{k-1}+\frac12(2k-1)(2k-3)\right)\right)\chi_{i}\\&
+\frac1{16}(i+m_{k}+m_{k-1}+2k-3)(i+\nu)(i-m_{k}-m_{k-1}-2k+1)(i-2-\nu)\chi_{i}\\&=
\frac1{16}(i-m_{k}-m_{k-1}-2k+3)(i-\nu)(i+m_{k}+m_{k-1}+2k-1)(i+2+\nu)\chi_{i},
\end{align*}
due to the identity
\begin{align*}
&(i-m_{k}-m_{k-1}-2k+3)(i-\nu)(i+m_{k}+m_{k-1}+2k-1)(i+2+\nu)\\
&-(i+m_{k}+m_{k-1}+2k-3)(i+\nu)(i-m_{k}-m_{k-1}-2k+1)(i-2-\nu)\\
&=8i^{3}-8i\left(m_{k}^{2}
+m_{k-1}^{2}+(2k-1)m_{k}+(2k-3)m_{k-1}+\frac12(2k-1)(2k-3)\right).
\end{align*}
Since $ (i-m_{k}-m_{k-1}-2k+3)(i-\nu)\ne0$ we obtain
$$
D^{-}\chi_{i+2}=\frac14(i+m_{k}+m_{k-1}+2k-1)(i+2+\nu)\chi_{i}
$$
that completes the induction.
\end{proof}

Lemma \ref{PrD-actions}, expansion (\ref{ExpanTilde}) and
relations (\ref{DDrelations}) effectively describe the action of
operators $D_{0},D_{1},D_{2},D_{3}$ in the space
$\mathcal{L}^{2}\left(\SO(2k+1),\SO(2k-1),\mu\right)$. Consider
the problem of finding all common eigenvectors $\psi_{D}$ of
operators $D_{0}^{2},D_{1},D_{2}$ and optionally $D_{3}$. It is
equivalent to the problem of finding all common eigenvectors of
operators $D_{0}^{2},D^{+}+D^{-}$ and optionally $D^{+}-D^{-}$ in
the space $\widetilde{V}_{\mathfrak{B}_{k}}(\lambda)$.

Eigenvectors for the operator $D_{0}^{2}$ are
$$
c_{+}\chi_{j}+c_{-}\chi_{-j},\,c_{\pm}\in\mathbb{C},\,j\in
L_{\nu},\,j\geqslant0
$$
with eigenvalues $-j^{2}$. Since
\begin{align*}
\left(D^{+}+D^{-}\right)\left(c_{+}\chi_{j}+c_{-}\chi_{-j}\right)&=
\frac14(j-m_{k}-m_{k-1}-2k+3)(j-\nu)\left(c_{+}\chi_{j+2}+c_{-}\chi_{-j-2}\right)\\
&+\frac14(j+m_{k}+m_{k-1}+2k-3)(j+\nu)\left(c_{+}\chi_{j-2}+c_{-}\chi_{-j+2}\right),
\end{align*}
the requirement
$$
\left(D^{+}+D^{-}\right)\left(c_{+}\chi_{j}+c_{-}\chi_{-j}\right)\sim
c_{+}\chi_{j}+c_{-}\chi_{-j}
$$
implies $(j-m_{k}-m_{k-1}-2k+3)(j-\nu)=0$ that leads to two cases:
$j=m_{k}-m_{k-1}$ and $j=m_{k}+m_{k-1}+2k-3$.

In the first case one gets
\begin{align*}
\left(D^{+}+D^{-}\right)&\left(c_{+}\chi_{m_{k}-m_{k-1}}+c_{-}\chi_{-m_{k}+m_{k-1}}\right)
\\&= (m_{k}-m_{k-1})\left(m_{k}+k-\frac32\right)
\left(c_{+}\chi_{m_{k}-m_{k-1}-2}+c_{-}\chi_{-m_{k}+m_{k-1}+2}\right)
\end{align*}
that implies one of three possibilities
\begin{enumerate}
\item $m_{k}-m_{k-1}=0$;
\item $m_{k}-m_{k-1}-2=-m_{k}+m_{k-1}$;
\item $m_{k}-m_{k-1}-2=0,\;c_{+}+c_{-}=0$.
\end{enumerate}
Thus we obtain the following eigenvectors:
\begin{enumerate}
\item $\left(D^{+}+D^{-}\right)\chi_{0}=0$ for $m_{k}-m_{k-1}=0$;
\item $\left(D^{+}+D^{-}\right)\left(\chi_{1}+\chi_{-1}\right)=
\left(m_{k}+k-\frac32\right)\left(\chi_{1}+\chi_{-1}\right)$ for
$m_{k}\in\mathbb{N},\,m_{k-1}=m_{k}-1$;
\item $\left(D^{+}+D^{-}\right)\left(\chi_{1}-\chi_{-1}\right)=
-\left(m_{k}+k-\frac32\right)\left(\chi_{1}-\chi_{-1}\right)$ for
$m_{k}\in\mathbb{N},\,m_{k-1}=m_{k}-1$;
\item $\left(D^{+}+D^{-}\right)\left(\chi_{2}-\chi_{-2}\right)=0,\,m_{k-1}=m_{k}-2,\,
m_{k}=2,3,\ldots$.
\end{enumerate}

In the second case one gets $ m_{k}+m_{k-1}+2k-3=j\leqslant
m_{k}-m_{k-1}$ that implies $0\leqslant m_{k-1}\leqslant\frac32-k$
and thus $k=1$ that contradicts to the assumption $k\geqslant 2$.

Using relations (\ref{DDrelations}) this consideration can be
summarized in the following proposition.
\begin{proposit}\label{eigenvectorsOdd}
For $n=2k,\,k\geqslant2$ there are four series of common
eigenvectors in
$\widetilde{V}_{\mathfrak{B}_{k}}(m_{k}\varepsilon_{k}+m_{k-1}\varepsilon_{k-1}),\;
m_{k},m_{k-1}\in\mathbb{Z}_{+}$ for the operators
$D_{0}^{2},D_{1},D_{2}$:
\begin{enumerate}
\item
$D_{0}^{2}\chi_{0}=D_{3}\chi_{0}=0,\,D_{1}\chi_{0}=D_{2}\chi_{0}=-m_{k}(m_{k}+2k-2)\chi_{0},\;
m_{k}=m_{k-1}$;
\item
$D_{0}^{2}(\chi_{1}+\chi_{-1})=-(\chi_{1}+\chi_{-1}),\;
D_{2}(\chi_{1}+\chi_{-1})=-m_{k}(m_{k}+2k-2)(\chi_{1}+\chi_{-1}),\\
D_{1}(\chi_{1}+\chi_{-1})=\left(-m_{k}^{2}-2(k-2)m_{k}+2k-3\right)(\chi_{1}+\chi_{-1}),\\
D_{3}(\chi_{1}+\chi_{-1})=\ii\left(m_{k}+k-\frac32\right)
(\chi_{1}-\chi_{-1}),\;m_{k-1}=m_{k}-1,m_{k}\in\mathbb{N}$
\item
$D_{0}^{2}(\chi_{1}-\chi_{-1})=-(\chi_{1}-\chi_{-1}),\;
D_{1}(\chi_{1}-\chi_{-1})=-m_{k}(m_{k}+2k-2)(\chi_{1}-\chi_{-1}),\\
D_{2}(\chi_{1}-\chi_{-1})=\left(-m_{k}^{2}-2(k-2)m_{k}+2k-3\right)(\chi_{1}-\chi_{-1}),\\
D_{3}(\chi_{1}-\chi_{-1})=-\ii\left(m_{k}+k-\frac32\right)
(\chi_{1}+\chi_{-1}),\;m_{k-1}=m_{k}-1,m_{k}\in\mathbb{N}$;
\item
$D_{0}^{2}(\chi_{2}-\chi_{-2})=-4(\chi_{2}-\chi_{-2}),\;
D_{3}(\chi_{2}-\chi_{-2})=-4\ii\left(m_{k}+k-\frac32\right)\chi_{0},\\
D_{1}(\chi_{2}-\chi_{-2})=D_{2}(\chi_{2}-\chi_{-2})=
\left(-m_{k}^{2}-2(k-2)m_{k}+2k-3\right)(\chi_{2}-\chi_{-2}),\\
m_{k-1}=m_{k}-2,m_{k}=2,3,4,\ldots$
\end{enumerate}
Only the first vector is also an eigenvector for the operator
$D_{3}$.

Multiplicities of corresponding eigenvalues in
$\mathcal{L}^{2}\left(\SO(n+1),\SO(n-1),\mu\right)$ are equal to
$\dim
V_{\mathfrak{B}_{k}}\left(m_{k}\varepsilon_{k}+m_{k-1}\varepsilon_{k-1}\right)$
and can be calculated in explicit form using (\ref{Weyl}).
\end{proposit}

Consider the case $k=1,n=2$. Now the group $K$ is trivial and
therefore
$\widetilde{V}_{\mathfrak{B}_{1}}(\lambda)=V_{\mathfrak{B}_{1}}(\lambda)$.
The algebra
$\mathfrak{B}_{1}=\mathfrak{so}(3,\mathbb{C})\cong\mathfrak{sl}(2,\mathbb{C})=\mathfrak{A}_{1}$
is spanned by elements $F_{11},F_{01},F_{10}$ with commutator
relations
$$
[F_{11},F_{01}]=-F_{01},\;[F_{11},F_{10}]=F_{10},\;[F_{10},F_{01}]=F_{11}.
$$
Its representation theory is well known: all its finite
dimensional irreducible modules are of the form
$$
V_{\mathfrak{B}_{1}}(m\varepsilon_{1})=V_{-m\varepsilon_{1}}\oplus
V_{-(m-1)\varepsilon_{1}}\oplus\ldots\oplus
V_{(m-1)\varepsilon_{1}}\oplus V_{m\varepsilon_{1}},
$$
where $m\in\mathbb{Z}_{+}\cup\left(\mathbb{Z}_{+}+\frac12\right)$,
all $V_{j\varepsilon_{1}}$ are one-dimensional weight subspaces
w.r.t.\ $\mathfrak{h}_{1}=\lspan(F_{11})$ and
$$
F_{10}:\;V_{j\varepsilon_{1}}\rightarrow
V_{(j+1)\varepsilon_{1}},\;\,j=-m,\ldots,m-1,\;
F_{01}:\;V_{j\varepsilon_{1}}\rightarrow
V_{(j-1)\varepsilon_{1}},\,j=-m+1,\ldots,m
$$
are bijections.

We shall consider only $m\in\mathbb{Z}_{+}$ since
$$
\mathcal{L}^{2}(\SO(3),\mu)=\bigoplus_{m\in\mathbb{Z}_{+}}
(2m+1)V_{\mathfrak{B}_{1}}(m\varepsilon_{1}).
$$
Thus there are additional weight subspaces in the module
$V_{\mathfrak{B}_{1}}(m\varepsilon_{1})$ w.r.t.\ expansion
(\ref{MainExpansion}) and the action of the algebra, generated by
the operators $D^{+}=\frac12F^{2}_{10},D^{-}=\frac12F^{2}_{01}$,
is not irreducible in $V_{\mathfrak{B}_{1}}(m\varepsilon_{1})$.

One can choose a base $\left(\chi_{j}\right)_{j=-m}^{m}$ in
$V_{\mathfrak{B}_{1}}(m\varepsilon_{1})$ such that
\begin{gather*}
\chi_{j}\in V_{j\varepsilon_{1}},\,F_{11}\chi_{j}=j\chi_{j},\,
F_{10}\chi_{j}=-\frac1{\sqrt{2}}\sqrt{(m-j)(m+j+1)}\chi_{j+1},\\
F_{01}\chi_{j}=-\frac1{\sqrt{2}}\sqrt{(m+j)(m-j+1)}\chi_{j-1},
\end{gather*}
where as above $\chi_{j}=0$ for $|j|>m$.

Eigenvectors for the operator $D_{0}^{2}=-F_{11}^{2}$ are
$$
c_{+}\chi_{j}+c_{-}\chi_{-j},\,c_{\pm}\in\mathbb{C},\,j=0,1,\ldots,m
$$
with eigenvalues $-j^{2}$. Since
\begin{align*}
\left(D^{+}+D^{-}\right)&\left(c_{+}\chi_{j}+c_{-}\chi_{-j}\right)\\&=
\frac14\sqrt{(m-j)(m+j+1)(m-j-1)(m+j+2)}
\left(c_{+}\chi_{j+2}+c_{-}\chi_{-j-2}\right)\\
&+\frac14\sqrt{(m+j)(m-j+1)(m+j-1)(m-j+2)}
\left(c_{+}\chi_{j-2}+c_{-}\chi_{-j+2}\right),
\end{align*}
the requirement
$$
\left(D^{+}+D^{-}\right)\left(c_{+}\chi_{j}+c_{-}\chi_{-j}\right)\sim
c_{+}\chi_{j}+c_{-}\chi_{-j}
$$
implies $(m-j)(m+j+1)(m-j-1)(m+j+2)=0$ that gives two cases: $j=m$
and $j=m-1$.

In the first case one gets
\begin{align*}
\left(D^{+}+D^{-}\right)&\left(c_{+}\chi_{m}+c_{-}\chi_{-m}\right)
=\frac12\sqrt{m(2m-1)}
\left(c_{+}\chi_{m-2}+c_{-}\chi_{m+2}\right)
\end{align*}
that implies one of three possibilities
\begin{enumerate}
\item $m=j=0$;
\item $m-2=-m$;
\item $m-2=0,\;c_{+}+c_{-}=0$.
\end{enumerate}
This gives the following eigenvectors:
\begin{enumerate}
\item $\left(D^{+}+D^{-}\right)\chi_{0}=0,\;m=0$;
\item $\left(D^{+}+D^{-}\right)\left(\chi_{1}+\chi_{-1}\right)=\frac12
\left(\chi_{1}+\chi_{-1}\right),\;m=1$;
\item $\left(D^{+}+D^{-}\right)\left(\chi_{1}-\chi_{-1}\right)=
-\frac12 \left(\chi_{1}-\chi_{-1}\right),\;m=1$;
\item $\left(D^{+}+D^{-}\right)\left(\chi_{2}-\chi_{-2}\right)=0,\,m=2$.
\end{enumerate}

It is easily seen that these eigenvectors corresponds to
eigenvectors from proposition \ref{eigenvectorsOdd} for
$m_{k}=m,m_{k-1}=0$.

In the second case it holds
\begin{align*}
\left(D^{+}+D^{-}\right)&\left(c_{+}\chi_{m-1}+c_{-}\chi_{-m+1}\right)
=\frac12\sqrt{3(2m-1)(m-1)}
\left(c_{+}\chi_{m-3}+c_{-}\chi_{m+3}\right)
\end{align*}
that implies one of three possibilities
\begin{enumerate}
\item $m=1,\,j=0$;
\item $m-3=-m+1$;
\item $m-3=0,\;c_{+}+c_{-}=0$.
\end{enumerate}
Thus one gets the following eigenvectors:
\begin{enumerate}
\item $\left(D^{+}+D^{-}\right)\chi_{0}=0,\;m=1$;
\item $\left(D^{+}+D^{-}\right)\left(\chi_{1}+\chi_{-1}\right)=\frac32
\left(\chi_{1}+\chi_{-1}\right),\;m=2$;
\item $\left(D^{+}+D^{-}\right)\left(\chi_{1}-\chi_{-1}\right)=
-\frac32 \left(\chi_{1}-\chi_{-1}\right),\;m=2$;
\item $\left(D^{+}+D^{-}\right)\left(\chi_{2}-\chi_{-2}\right)=0,\,m=3$.
\end{enumerate}

Since it holds
$\left.\widetilde{C}\right|_{\widetilde{V}_{\mathfrak{B}_{1}}(m\varepsilon_{1})}=m(m+1)\id$
and relations (\ref{DDrelations}) are valid also in the case $k=1$
one gets the following proposition.
\begin{proposit}\label{eigenvectorsk=1}
There are eight common eigenvectors in
$V_{\mathfrak{B}_{1}}(m\varepsilon_{1})$ for the operators
$D_{0}^{2},D_{1},D_{2}$:
\begin{enumerate}
\item $D_{0}^{2}\chi_{0}=D_{1}\chi_{0}=D_{2}\chi_{0}=D_{3}\chi_{0}=0,\,m=0$;
\item $D_{0}^{2}\chi_{0}=D_{3}\chi_{0}=0,\,D_{1}\chi_{0}=D_{2}\chi_{0}=-\chi_{0},\,m=1$;
\item $D_{0}^{2}(\chi_{1}+\chi_{-1})=D_{2}(\chi_{1}+\chi_{-1})=-(\chi_{1}+\chi_{-1}),\,
D_{1}(\chi_{1}+\chi_{-1})=0,\\
D_{3}(\chi_{1}+\chi_{-1})=\frac{\ii}2(\chi_{1}-\chi_{-1}),\;m=1;$
\item $D_{0}^{2}(\chi_{1}-\chi_{-1})=D_{1}(\chi_{1}-\chi_{-1})=-(\chi_{1}-\chi_{-1}),\,
D_{2}(\chi_{1}-\chi_{-1})=0,\\
D_{3}(\chi_{1}-\chi_{-1})=-\frac{\ii}2(\chi_{1}+\chi_{-1}),\;m=1;$
\item $D_{0}^{2}(\chi_{2}-\chi_{-2})=-4(\chi_{2}-\chi_{-2}),\,D_{1}(\chi_{2}-\chi_{-2})=
D_{2}(\chi_{2}-\chi_{-2})=-(\chi_{2}-\chi_{-2}),\\
D_{3}(\chi_{2}-\chi_{-2})=-\sqrt{6}\ii\chi_{0},\;m=2;$
\item $D_{0}^{2}(\chi_{1}+\chi_{-1})=D_{1}(\chi_{1}+\chi_{-1})=-(\chi_{1}+\chi_{-1}),\,
D_{2}(\chi_{1}+\chi_{-1})=-4(\chi_{1}+\chi_{-1}),\\
D_{3}(\chi_{1}+\chi_{-1})=\frac32\ii(\chi_{1}-\chi_{-1}),\;m=2;$
\item $D_{0}^{2}(\chi_{1}-\chi_{-1})=D_{2}(\chi_{1}-\chi_{-1})=-(\chi_{1}-\chi_{-1}),\,
D_{1}(\chi_{1}-\chi_{-1})=-4(\chi_{1}-\chi_{-1}),\\
D_{3}(\chi_{1}-\chi_{-1})=-\frac32\ii(\chi_{1}+\chi_{-1}),\;m=2;$
\item $D_{0}^{2}(\chi_{2}-\chi_{-2})=D_{1}(\chi_{2}-\chi_{-2})=
D_{2}(\chi_{2}-\chi_{-2})=-4(\chi_{2}-\chi_{-2}),\\
D_{3}(\chi_{2}-\chi_{-2})=-\sqrt{30}\ii\chi_{0},\;m=3.$
\end{enumerate}
Only the first and the second vectors are also eigenvectors for
the operator $D_{3}$.

Multiplicities of corresponding eigenvalues in
$\mathcal{L}^{2}\left(\SO(3),\mu\right)$ are $2m+1$.
\end{proposit}

\subsection{The case $n=2k-1$}\label{n=2k-1}

Here we use notations from appendix \ref{DCaseAp}. The algebra
$\mathfrak{D}_{k}$ is considered there as a subalgebra of
$\mathfrak{B}_{k}$. Therefore one can easily obtain analogs of
formulas (\ref{DPsi}) simply by deleting the terms $F_{k0}$ and
$F_{0k}$:
\begin{align*}
D_{0}&=-\ii F_{kk},\;D_{1}=
\frac12\sum_{j=1}^{k-1}\left\{F_{-kj}+F_{kj},
F_{k,-j}+F_{-k,-j}\right\},\notag\\ D_{2}&=
\frac12\sum_{j=1}^{k-1}\left\{F_{-kj}-F_{kj},
F_{k,-j}-F_{-k,-j}\right\},\; D_{3}=
\ii\sum_{j=1}^{k-1}\left(F_{kj}F_{k,-j}-
F_{-kj}F_{-k,-j}\right).\notag
\end{align*}

Let the space $\mathcal{R}_{\ell,i}$ equals
$V_{\mathfrak{D}_{k}}(\lambda)$ for a highest weight
(\ref{HigestWeightD}), where
$m_{i}\in\mathbb{Z}_{+},\,i\geqslant2,m_{1}\in\mathbb{Z}$, and
$\widetilde{V}_{\mathfrak{D}_{k}}(\lambda)$ be a subspace of
$V_{\mathfrak{D}_{k}}(\lambda)$ annulled by the subalgebra
$\mathfrak{k}^{\mathbb{C}}\cong\mathfrak{D}_{k-1}$. Reasoning as
above in the case $n=2k$, one gets that
$\widetilde{V}_{\mathfrak{D}_{k}}(\lambda)$ is nontrivial iff
\begin{equation}\label{LambdaCon}
\lambda=m_{k}\varepsilon_{k}+m_{k-1}\varepsilon_{k-1},\;m_{k}\geqslant|m_{k-1}|,\,
m_{k}\in\mathbb{Z}_{+},\,m_{k-1}\in\mathbb{Z}'_{k},
\end{equation}
where $\mathbb{Z}'_{k}=\mathbb{Z}_{+}$ for $k\geqslant3$ and
$\mathbb{Z}'_{2}=\mathbb{Z}$. In this case one has
\begin{equation}\label{DIM2}
\dim\widetilde{V}_{\mathfrak{D}_{k}}(\lambda)=m_{k}-|m_{k-1}|+1.
\end{equation}

Below in the present subsection we suppose that condition
(\ref{LambdaCon}) is valid. This leads to the expansion:
\begin{align*}
\mathcal{L}^{2}&\left(\SO(2k),\SO(2k-2),\mu\right)\\ &=
\bigoplus_{\genfrac{}{}{0pt}{0}{m_{k}\geqslant
|m_{k-1}|}{m_{k}\in\mathbb{Z}_{+},\,m_{k-1}\in\mathbb{Z}'_{k}}}
(m_{k}-|m_{k-1}|+1)V_{\mathfrak{D}_{k}}
\left(m_{k}\varepsilon_{k}+m_{k-1}\varepsilon_{k-1}\right)
\end{align*}
of the left $\SO(2k)$-space
$\mathcal{L}^{2}\left(\SO(2k),\SO(2k-2),\mu\right)$ and to the
expansion:
\begin{align}\label{ExpanTildeD}
\mathcal{L}^{2}&\left(\SO(2k),\SO(2k-2),\mu\right)\\&=
\bigoplus_{\genfrac{}{}{0pt}{0}{m_{k}\geqslant
|m_{k-1}|}{m_{k}\in\mathbb{Z}_{+},\,m_{k-1}\in\mathbb{Z}'_{k}}}\left(\dim
V_{\mathfrak{D}_{k}}\left(m_{k}\varepsilon_{k}+m_{k-1}\varepsilon_{k-1}\right)\right)
\widetilde{V}_{\mathfrak{D}_{k}}\left(m_{k}\varepsilon_{k}+m_{k-1}
\varepsilon_{k-1}\right),\notag
\end{align}
of the same space as a
$\Diff_{\SO(2k)}(\SO(2k)/\SO(2k-2))$-module, where the dimension
$\dim
V_{\mathfrak{D}_{k}}\left(m_{k}\varepsilon_{k}+m_{k-1}\varepsilon_{k-1}\right)$
is given by (\ref{Weyl}).

Now let
\begin{align*}
D^{+}&:=\sum_{j=1}^{k-1}F_{kj}F_{k,-j},\,
D^{-}:=\sum_{j=1}^{k-1}F_{-kj}F_{-k,-j},\\
\widetilde{C}&:=\left.C\right|_{\mathcal{L}^{2}\left(\SO(2k),\SO(2k-2),\mu\right)}
=F_{kk}^{2}+
\sum_{j=1}^{k-1}\left(\{F_{kj},F_{jk}\}+\{F_{k,-j},F_{-jk}\}\right)
\end{align*}
be operators from $\Diff_{\SO(2k)}(\SO(2k)/\SO(2k-2))$, where $C$
is the universal Casimir operator (\ref{CasimirD}).

Formulas (\ref{DDrelations}) and (\ref{FDComm}) are valid without
any modification and formula (\ref{DDComm}) becomes
$$
[D^{+},D^{-}]=-\frac12F_{kk}^{3}+\frac12\widetilde{C}F_{kk}+
(k-1)(k-2)F_{kk}.
$$
Now
\begin{equation*}\label{CasimirEigenDCase}
\left.\widetilde{C}\right|_{\widetilde{V}_{\mathfrak{D}_{k}}(\lambda)}=
\left(\left(m_{k}+k-1\right)^{2}+\left(m_{k-1}+k-2\right)^{2}
-\left(k-1\right)^{2}-\left(k-2\right)^{2}\right)\id.
\end{equation*}
From \cite{Mol1} it follows\footnote{See also appendix \ref{appD}
for a proof independent from \cite{Mol1}.} that
\begin{equation}\label{MainExpansion1}
\widetilde{V}_{\mathfrak{D}_{k}}(\lambda)=V_{-\nu\varepsilon_{k}}\oplus
V_{-(\nu-2)\varepsilon_{k}}\oplus\ldots\oplus
V_{(\nu-2)\varepsilon_{k}}\oplus V_{\nu\varepsilon_{k}},
\end{equation}
where $\nu=m_{k}-|m_{k-1}|$, all summands are one-dimensional
weight spaces w.r.t.\ the Cartan subalgebra
$\mathfrak{h}_{k}\subset\mathfrak{D}_{k}$ and the algebra,
generated by the operators $D^{+},D^{-}$, acts in
$\widetilde{V}_{\mathfrak{D}_{k}}(\lambda)$ in an irreducible way.

Again we shall simplify formulas for this action w.r.t.\
\cite{Mol1} using another base. The next lemma can be proved
completely similar to the proof of lemma \ref{PrD-actions}.

\begin{Lem}\label{PrD-actionsDCase}
Let
$\nu:=m_{k}-|m_{k-1}|,\;L_{\nu}:=(-\nu,-\nu+2,\ldots,\nu-2,\nu)$.
There is a base $\left(\chi_{j}\right)_{j\in L_{\nu}}$ in
$\widetilde{V}_{\mathfrak{D}_{k}}(\lambda)$ such that
\begin{align*}
F_{kk}\chi_{j}=j\chi_{j}&,\;
D^{+}\chi_{j}=\frac14(j-m_{k}-|m_{k-1}|-2k+4)(j-\nu)\chi_{j+2},\\
D^{-}\chi_{j}&=\frac14(j+m_{k}+|m_{k-1}|+2k-4)(j+\nu)\chi_{j-2},
\end{align*}
where $\chi_{j}=0$ if $j\not\in L_{\nu}$.
\end{Lem}

Arguing as in the $\mathfrak{B}_{k}$-case one gets the following
proposition.

\begin{proposit}\label{eigenvectorsEven}
For $n=2k-1,\,k\geqslant2$ there are four series of common
eigenvectors in
$\widetilde{V}_{\mathfrak{D}_{k}}(m_{k}\varepsilon_{k}+|m_{k-1}|\varepsilon_{k-1}),\;
m_{k}\in\mathbb{Z}_{+},\,m_{k-1}\in\mathbb{Z}'_{k}$ for the
operators $D_{0}^{2},D_{1},D_{2}$:
\begin{enumerate}
\item
$D_{0}^{2}\chi_{0}=D_{3}\chi_{0}=0,\,D_{1}\chi_{0}=D_{2}\chi_{0}=-m_{k}(m_{k}+2k-3)\chi_{0},\;
m_{k}=|m_{k-1}|$;
\item
$D_{0}^{2}(\chi_{1}+\chi_{-1})=-(\chi_{1}+\chi_{-1}),\;
D_{2}(\chi_{1}+\chi_{-1})=-m_{k}(m_{k}+2k-3)(\chi_{1}+\chi_{-1}),\\
D_{1}(\chi_{1}+\chi_{-1})=\left(-m_{k}^{2}+(5-2k)m_{k}+2k-4\right)(\chi_{1}+\chi_{-1}),\\
D_{3}(\chi_{1}+\chi_{-1})=\ii\left(m_{k}+k-2\right)
(\chi_{1}-\chi_{-1}),\;|m_{k-1}|=m_{k}-1,m_{k}\in\mathbb{N}$
\item
$D_{0}^{2}(\chi_{1}-\chi_{-1})=-(\chi_{1}-\chi_{-1}),\;
D_{1}(\chi_{1}-\chi_{-1})=-m_{k}(m_{k}+2k-3)(\chi_{1}-\chi_{-1}),\\
D_{2}(\chi_{1}-\chi_{-1})=\left(-m_{k}^{2}+(5-2k)m_{k}+2k-4\right)(\chi_{1}-\chi_{-1}),\\
D_{3}(\chi_{1}-\chi_{-1})=-\ii\left(m_{k}+k-2\right)
(\chi_{1}+\chi_{-1}),\;|m_{k-1}|=m_{k}-1,m_{k}\in\mathbb{N}$;
\item
$D_{0}^{2}(\chi_{2}-\chi_{-2})=-4(\chi_{2}-\chi_{-2}),\;
D_{3}(\chi_{2}-\chi_{-2})=-4\ii\left(m_{k}+k-2\right)\chi_{0},\\
D_{1}(\chi_{2}-\chi_{-2})=D_{2}(\chi_{2}-\chi_{-2})=
\left(-m_{k}^{2}+(5-2k)m_{k}+2k-4\right)(\chi_{2}-\chi_{-2}),\\
|m_{k-1}|=m_{k}-2,m_{k}=2,3,4,\ldots$
\end{enumerate}
Only the first vector is also an eigenvector for the operator
$D_{3}$.

Multiplicities of corresponding eigenvalues in
$\mathcal{L}^{2}\left(\SO(n+1),\SO(n-1),\mu\right)$ are equal to
$\dim
V_{\mathfrak{D}_{k}}\left(m_{k}\varepsilon_{k}+m_{k-1}\varepsilon_{k-1}\right)$
and can be calculated in explicit form using (\ref{Weyl}).
\end{proposit}
\begin{Rem}
For $k=2$ a value $m_{k-1}=m_{1}$ can has an arbitrary sign and
one gets eight common eigenvectors found in \cite{ShchStep}.
\end{Rem}
\begin{Rem}\label{Rem2}
Results of propositions \ref{eigenvectorsOdd},
\ref{eigenvectorsk=1} and \ref{eigenvectorsEven} correspond to
proposition \ref{aprioriInf} and are even more restrictive.
Indeed, if
$\psi_{D}\in\mathcal{L}^{2}\left(\SO(n+1),\SO(n-1),\mu\right)$ is
an eigenfunction for operators $D_{0}^{2},D_{1},D_{2}$ and
$D_{3}$, then
$D_{0}\psi_{D}=D_{3}\psi_{D}=0,\,D_{1}\psi_{D}=D_{2}\psi_{D}$.
\end{Rem}

\section{Scalar spectral equations and some energy levels for the two-body problem in
${\bf S}^{n}$} \label{SpectralEquationTBP}
\markright{\ref{SpectralEquationTBP} Spectral equations for the
two-body problem}

Here we shall consider the spectral problem (\ref{StatSchrEq}),
where the operator $H$ is defined in (\ref{TBHamR1}) and
$\psi_{D}$ is one of common eigenfunctions for operators
$D_{0}^{2},D_{1},D_{2}$ and optionally $D_{3}$. In this section
$m$ denotes reduced mass (\ref{RedMass}) and integers $m_{k}$
correspond to highest weights in $\mathfrak{so}(n)$-modules.

Let
$D_{0}^{2}\psi_{D}=\delta_{0}\psi_{D},\,D_{i}\psi_{D}=\delta_{i}\psi_{D},\,i=1,2$.
In accordance with remark \ref{Rem2} there are two main cases:
\begin{enumerate}
\item
$D_{3}\psi_{D}=0,\,\delta_{0}=0,\delta_{1}=\delta_{2}$, particle
masses are arbitrary;
\item $D_{3}\psi_{D}\not\sim\psi_{D}$, particle masses are equal.
\end{enumerate}
In the first case
$$
\left(CD_{1}+AD_{2}+2BD_{3}\right)\psi_{D}=\delta_{1}(C+A)\psi_{D}=
\frac{(1+r^{2})^{2}}{4mR^{2}r^{2}}\delta_{1}\psi_{D}.
$$
In the second case
$$
A=\frac{1+r^{2}}{4mR^{2}r^{2}},\,B\equiv0,\,C=\frac{1+r^{2}}{4mR^{2}}.
$$
In all cases one gets the following spectral equation for the
function $f(r)$
\begin{align}\begin{split}
f''+\frac{n-1+(3-n)r^{2}}{(1+r^{2})r}f'&+\frac8{(1+r^{2})^{2}}\left(mR^{2}(E-V(r))-
\frac{a}{r^{2}}-b-cr^{2}\right)f=0,\\ a,b,c\geqslant0,\;
0&<r<\infty.
\end{split}\label{SpectralEqGenForm}\end{align}
where coefficients $a,b,c$ are described below.

For eigenfunctions $\psi_{D}$ classified in proposition
\ref{eigenvectorsOdd} ($n=2k,k=2,3,\ldots$) one has
\begin{enumerate}
\item
$a=c=m_{k}(m_{k}+2k-2)/8,\,b=2a,\,m_{k}\in\mathbb{Z}_{+}$, masses
are arbitrary;
\item
$a=m_{k}(m_{k}+2k-2)/8,\,b=(m_{k}^{2}+(2k-3)m_{k}-k+2)/4,\,c=(m_{k}^{2}+2(k-2)m_{k}-2k+3)/8,\,
m_{k}\in\mathbb{N}$, masses are equal;
\item
$a=(m_{k}^{2}+2(k-2)m_{k}-2k+3)/8,\,b=(m_{k}^{2}+(2k-3)m_{k}-k+2)/4,\,c=m_{k}(m_{k}+2k-2)/8,\,
m_{k}\in\mathbb{N}$, masses are equal;
\item
$a=c=(m_{k}^{2}+2(k-2)m_{k}-2k+3)/8,\,b=(m_{k}^{2}+2(k-2)m_{k}-2k+5)/4,\,
m_{k}=2,3,\ldots$, masses are equal.
\end{enumerate}

Proposition \ref{eigenvectorsk=1} ($n=2$) gives the following
values for $a,b,c$:
\begin{enumerate}
\item
$a=c=b=0$, masses are arbitrary;
\item
$a=c=1/8,b=1/4$, masses are arbitrary;
\item
$a=1/8,b=1/4,c=0$, masses are equal;
\item
$a=0,b=1/4,c=1/8$, masses are equal;
\item
$a=c=1/8,b=3/4$, masses are equal;
\item
$a=1/2,b=3/4,c=1/8$, masses are equal;
\item
$a=1/8,b=3/4,c=1/2$, masses are equal;
\item
$a=c=1/2,b=3/2$, masses are equal.
\end{enumerate}

Finally, proposition \ref{eigenvectorsEven} corresponds to the
following cases ($n=2k-1,k=2,3,\ldots$)
\begin{enumerate}
\item
$a=c=m_{k}(m_{k}+2k-3)/8,\,b=2a,\,m_{k}\in\mathbb{Z}_{+}$, masses
are arbitrary;
\item
$a=m_{k}(m_{k}+2k-3)/8,\,b=(m_{k}^{2}+(2k-4)m_{k}-k+\dfrac52)/4,\,
c=(m_{k}^{2}+(2k-5)m_{k}-2k+4)/8,\,m_{k}\in\mathbb{N}$, masses are
equal;
\item
$a=(m_{k}^{2}+(2k-5)m_{k}-2k+4)/8,\,b=(m_{k}^{2}+(2k-4)m_{k}-k+\dfrac52)/4,\,
c=m_{k}(m_{k}+2k-3)/8,\,m_{k}\in\mathbb{N}$, masses are equal;
\item
$a=c=(m_{k}^{2}+(2k-5)m_{k}-2k+4)/8,\,b=(m_{k}^{2}+(2k-5)m_{k}-2k+6)/4,\,
m_{k}=2,3,\ldots$, masses are equal.
\end{enumerate}
We shall consider equation (\ref{SpectralEqGenForm}) for the
Coulomb and oscillator potentials.

\subsection{Coulomb potential}
For the Coulomb potential
\begin{equation}\label{CoulombPotential}
V_{c}=-\frac{\gamma}{R}\cot\frac{\rho}R=\frac{\gamma}{2R}\left(r-\frac1r\right),\;\gamma>0
\end{equation}
theorems \ref{TwoBodyHamComTh} and \ref{MilatTheorem} imply the
self-adjointness of the two-body Hamiltonian $H_{V_{c}}$ with its
domain defined by (\ref{MilatDomain}), where $V_{1}=0$ for $0<r<1$
and $V_{1}=V_{c}$ for $1\leqslant r<\infty$.

Equation (\ref{SpectralEqGenForm}) for $V=V_{c}$ is the Fuchsian
differential equation (see appendix \ref{AppendixD}) with four
singular points $r=0,\pm\ii,\infty$ and corresponding
characteristic exponents:
\begin{align}\label{ChExp}
&\rho^{(0)}_{\pm}=\frac12\left(2-n\pm\sqrt{(n-2)^{2}+32a}\right),\,
\rho^{(\infty)}_{\pm}=\frac12\left(2-n\pm\sqrt{(n-2)^{2}+32c}\right),\notag\\
&\rho^{(\ii)}_{\pm}=\frac12\left(n-1\pm\sqrt{\left(n-1\right)^{2}+8\left(mER^{2}-\ii
mR\gamma+a-b+c\right)}\right),\\
&\rho^{(-\ii)}_{\pm}=\frac12\left(n-1\pm\sqrt{\left(n-1\right)^{2}+8\left(mER^{2}+\ii
mR\gamma+a-b+c\right)}\right).\notag
\end{align}
Here and below we suppose that a square root for a positive number
is positive; for other numbers it is an arbitrary root.

The requirement $f(r)\psi_{D}\in\Dom\left(H_{V_{c}}\right)$
restricts asymptotics of $f(r)$ near singular points $r=0$ and
$r=\infty$. Let $f(r)\sim r^{\rho^{(0)}}$ as $r\to+0$ and
$f(r)\sim r^{-\rho^{(\infty)}}$ as $r\to+\infty$. We shall show
that $f(r)\psi_{D}\in\Dom\left(H_{V_{c}}\right)$ iff
$\rho^{(0)}=\rho^{(0)}_{+}$ and
$\rho^{(\infty)}=\rho^{(\infty)}_{+}$.

The inclusion
$$
f\in\mathcal{L}^{2}\left(\mathbb{R}_{+},\frac{r^{n-1}dr}{(1+r^{2})^{n}}\right)
$$
evidently implies $\rho^{(0)}>-n/2,\,\rho^{(\infty)}>-n/2$. On the
other hand one can easily see that the inequality $a\geqslant1/8$
leads to $\rho^{(0)}_{-}\leqslant-n/2$ and the inequality
$c\geqslant1/8$ leads to $\rho^{(\infty)}_{-}\leqslant-n/2$.

From the consideration above it follows that if $a<1/8\;(c<1/8)$
then $a=0\;(c=0)$. For $a=0$ the inequality
$\rho^{(0)}_{-}=2-n>-n/2$ implies $n<4$.

For $a=0,\,n=3$ the asymptotic $f(r)\sim r^{\rho^{(0)}_{-}}=1/r$
means that $\laplace(f\psi_{D})\sim\delta(0)$ as $r\to0$ that
contradicts to
\begin{equation}\label{LaplaceInclusion}
\laplace(f\psi_{D})\in\mathcal{L}^{2}_{\loc}\left({\bf
S}^{n}\times{\bf S}^{n},\chi\times\chi\right),
\end{equation}
see theorem \ref{MilatTheorem}.

For the case $a=0,\,n=2$ it holds
$\rho^{(0)}_{+}=\rho^{(0)}_{-}=0$ and the theory of Fuchsian
differential equations \cite{CodLev}, \cite{Go} implies that
canonical asymptotics of a solution for (\ref{SpectralEqGenForm})
near $r=0$ are $1$ and $\log r$. The latter asymptotic again leads
to $\laplace(f\psi_{D})\sim\delta(0)$ as $r\to0$ that again
contradicts to (\ref{LaplaceInclusion}).

Thus in all cases it should be $f(r)\sim r^{\rho^{(0)}_{+}}$ as
$r\to 0$. Reasoning in the similar way one gets also in all cases
the asymptotic $f(r)\sim r^{-\rho^{(\infty)}_{+}}$ as $r\to
+\infty$.

Consider the problem of reducing equation
(\ref{SpectralEqGenForm}) with potential (\ref{CoulombPotential})
to the hypergeometric equation via reducing
(\ref{SpectralEqGenForm}) to the Heun equation by transformations
(\ref{Mobius}), (\ref{Homotopy}) and then using theorem
\ref{HeunRedTh}.

Singular points of equation (\ref{SpectralEqGenForm}) form a
harmonic quadruple (see appendix \ref{AppendixD}). Therefore, one
can use only the first case of theorem \ref{HeunRedTh}. Move
singular points $(0,\pm\ii,\infty)$ of equation
(\ref{SpectralEqGenForm}) to the quadruple $(0,1,2,\infty)$ by a
fractional linear transformation $t=\tau(r)$ of independent
variable.

Since the order of singular points on a circle or on a line is
conserved by such transformation only two possibilities can occur.
The first one corresponds to the map of the unordered pair
$(\pm\ii)$ into the unordered pair $(0,2)$. The second one
corresponds to the map of the unordered pair $(0,\infty)$ into the
unordered pair $(0,2)$.

Then one can reduce the transformed equation to the Heun one by a
substitution of the form (\ref{Homotopy}). One of requirements of
the first case of theorem \ref{HeunRedTh} is the equality of
characteristic exponents at points $0$ and $2$. In terms of
characteristic exponents (\ref{ChExp}) it means that either
$|\rho^{(\ii)}_{+}-\rho^{(\ii)}_{-}|=|\rho^{(-\ii)}_{+}-\rho^{(-\ii)}_{-}|$
or
$|\rho^{(0)}_{+}-\rho^{(0)}_{-}|=|\rho^{(\infty)}_{+}-\rho^{(\infty)}_{-}|$.
The first possibility can not occur for a nontrivial $\gamma$.
Therefore, not loosing generality, one can consider the map
$$
t=\tau(r):=\frac{2r}{r+\ii},\quad\tau:\;(-\ii,0,\ii,\infty)\rightarrow(\infty,0,1,2).
$$

This map transforms equation (\ref{SpectralEqGenForm}) with
potential (\ref{CoulombPotential}) into the equation
\begin{equation}\label{PreHeun}
f_{tt}(t)+\mathcal{A}(t)f_{t}(t)-\mathcal{B}(t)f(t)=0,\;
|t-1|=1,\,\im t<0,
\end{equation}
where
\begin{gather*}
\mathcal{A}(t)=\frac{nt^{2}-2nt+2n-2}{t(t-1)(t-2)},\\
\mathcal{B}(t)=2\frac{m\left(ER^{2}t^{2}(t-2)^{2}+ R\gamma\ii
t(t-2)(t^{2}-2t+2)\right)
+a(t-2)^{4}-bt^{2}(t-2)^{2}+ct^{4}}{t^{2}(t-1)^{2}(t-2)^{2}}.
\end{gather*}
The substitution
$$
f(t)=t^{\rho^{(0)}_{+}}(t-1)^{\rho^{(\ii)}_{+}}(t-2)^{\rho^{(\infty)}_{+}}w(t)
$$
transforms (\ref{PreHeun}) into Heun equation (\ref{HeunEq}) with
the parameter $\gamma'$ instead of $\gamma$, where
\begin{gather*}
\alpha=\rho^{(0)}_{+}+\rho^{(\ii)}_{+}
+\rho^{(\infty)}_{+}+\rho^{(-\ii)}_{+},\,\beta=\rho^{(0)}_{+}+\rho^{(\ii)}_{+}
+\rho^{(\infty)}_{+}+\rho^{(-\ii)}_{-},\,d=2,\\
\gamma'=1-\rho^{(0)}_{-}+\rho^{(0)}_{+},\,\delta=1-\rho^{(\ii)}_{-}+\rho^{(\ii)}_{+},\,
\varepsilon=1-\rho^{(\infty)}_{-}+\rho^{(\infty)}_{+}.
\end{gather*}
Here
$t^{\rho^{(0)}_{+}}(t-1)^{\rho^{(\ii)}_{+}}(t-2)^{\rho^{(\infty)}_{+}}$
means the function holomorphic on
$\mathbb{C}\backslash(-\infty,2]$ and real for real $t>2$.
Restrictions on asymptotics of the function $f$ near the points
$r=0,\infty$ are equivalent to the boundedness of the function
$w(t)$ near the points $t=0,2$.

Obviously, the accessory parameter $q$ can be found as
\begin{align}\label{qCalcul}
q&=-2\lim_{t\to
0}t\left(-\mathcal{B}(t)+\left(\frac{\rho^{(0)}_{+}}{t}+
\frac{\rho^{(\ii)}_{+}}{t-1}+\frac{\rho^{(\infty)}_{+}}{t-2}
\right)\mathcal{A}(t)+\frac{\rho^{(0)}_{+}(\rho^{(0)}_{+}-1)}{t^{2}}+
\frac{2\rho^{(0)}_{+}\rho^{(\ii)}_{+}}{t(t-1)}\right.\notag\\
&+\left.\frac{2\rho^{(0)}_{+}\rho^{(\infty)}_{+}}{t(t-2)}\right)=
4\rho^{(0)}_{+}\rho^{(\ii)}_{+}+
2\rho^{(0)}_{+}\rho^{(\infty)}_{+}-(n-3)\rho^{(0)}_{+}+(n-1)
(2\rho^{(\ii)}_{+}+\rho^{(\infty)}_{+})\notag\\&-4mR\gamma\ii+16a.
\end{align}

Theorem \ref{HeunRedTh} implies that this Heun equation can be
transformed into the hypergeometric equation by a rational change
of independent variable $t\to z:\; z=P(r)$, where $P$ is a
rational function, iff
\begin{gather}\label{Gamma=Epsilon}
\gamma'=\varepsilon,\\ \label{q=AlphaBeta} \alpha\beta-q=0.
\end{gather}

Equation (\ref{Gamma=Epsilon}) is equivalent to
\begin{equation}\label{FirstCondit}
a=c.
\end{equation}

Using the equalities
\begin{align*}
\alpha&=\rho^{(0)}_{+}+\rho^{(\ii)}_{+}
+\rho^{(\infty)}_{+}+\frac12\left(n-1+
\sqrt{\left(n-1\right)^{2}+8\left(mER^{2}+\ii
mR\gamma+a-b+c\right)}\right),\\
\beta&=\rho^{(0)}_{+}+\rho^{(\ii)}_{+}
+\rho^{(\infty)}_{+}+\frac12\left(n-1-
\sqrt{\left(n-1\right)^{2}+8\left(mER^{2}+\ii
mR\gamma+a-b+c\right)}\right),
\end{align*}
one can rewrite equation (\ref{q=AlphaBeta}) as
\begin{align}\begin{split}
&\left(\rho^{(0)}_{+}+\rho^{(\ii)}_{+}
+\rho^{(\infty)}_{+}+\frac12(n-1)\right)^{2}-\frac14\left((n-1)^{2}
+8\left(mER^{2}+\ii mR\gamma+a-b+c\right)\right)\\&-
4\rho^{(0)}_{+}\rho^{(\ii)}_{+}-
2\rho^{(0)}_{+}\rho^{(\infty)}_{+}+(n-3)\rho^{(0)}_{+}-(n-1)
(2\rho^{(\ii)}_{+}+\rho^{(\infty)}_{+})+4mR\gamma\ii-16a\\&=\left(\rho^{(0)}_{+}\right)^{2}+
\left(\rho^{(\ii)}_{+}\right)^{2}+\left(\rho^{(\infty)}_{+}\right)^{2}+
2\rho^{(\ii)}_{+}\left(\rho^{(\infty)}_{+}-\rho^{(0)}_{+}\right)+
\left(2n-4\right)\rho^{(0)}_{+}\\&
-\left(n-1\right)\rho^{(\ii)}_{+}
+2mR\gamma\ii-2mER^{2}-18a+2b-2c=0.
\end{split}\label{EqInter1}
\end{align}

Excluding squares of values $\rho^{(0)}_{+},\,\rho^{(\ii)}_{+},
\,\rho^{(\infty)}_{+}$ from (\ref{EqInter1}) with the help of
obvious equations
\begin{gather*}
\left(\rho^{(0)}_{+}\right)^{2}+(n-2)\rho^{(0)}_{+}-8a=0,\\
\left(\rho^{(\ii)}_{+}\right)^{2}-(n-1)\rho^{(\ii)}_{+}-2mR\left(RE-
\gamma\ii\right)-2(a-b+c)=0,\\
\left(\rho^{(\infty)}_{+}\right)^{2}+(n-2)\rho^{(\infty)}_{+}-8c=0
\end{gather*}
for characteristic exponents, one gets
\begin{equation*}\label{EqInter2}
\left(2\rho^{(\ii)}_{+}-n+2\right)\left(\rho^{(\infty)}_{+}-\rho^{(0)}_{+}
\right)+8(c-a)=0.
\end{equation*}
For $a=c$ it holds $\rho^{(\infty)}_{+}=\rho^{(0)}_{+}$ and thus
equation (\ref{q=AlphaBeta}) is a consequence of
(\ref{FirstCondit}).

From now till the end of the present subsection we suppose that
$a=c$. This condition corresponds to cases 1,4 of proposition
\ref{eigenvectorsOdd}, cases 1,2,5,8 of proposition
\ref{eigenvectorsk=1}, and cases 1,4 of proposition
\ref{eigenvectorsEven}.

The fist case of theorem \ref{HeunRedTh} implies then that the
function $w$ w.r.t. a new independent variable
\begin{equation}\label{zt}
z:=1-(t-1)^{2}=t(2-t)
\end{equation}
satisfies the hypergeometric equation:
\begin{equation}\label{HypEqTilde}
z(1-z)w''(z)+(\widetilde{\gamma}-(\widetilde{\alpha}+\widetilde{\beta}+1)z)w'(z)-
\widetilde{\alpha}\widetilde{\beta} w(z)=0.
\end{equation}

The correspondence between characteristic exponents of the Heun
and the hypergeometric equations connected by (\ref{zt}) implies
\begin{align*}
\widetilde{\gamma}&=\gamma'=1+\sqrt{(n-2)^{2}+32a}\in\mathbb{R},\,
\widetilde{\alpha}=\frac12\alpha=\frac12+\frac12\sqrt{(n-2)^{2}+32a}+\frac14(s+\bar
s)\in\mathbb{R},\\
\widetilde{\beta}&=\frac12\beta=\frac12+\frac12\sqrt{(n-2)^{2}+32a}+\frac14(-s+\bar
s)\notin\mathbb{R},
\end{align*}
where $s=\sqrt{\left(n-1\right)^{2}+8\left(mER^{2}+\ii
mR\gamma+2a-b\right)}$.

Since $$z-1=-\left(\frac{r-i}{r+i}\right)^{2},$$ the half-line
$[0,\infty]$ on the $r$-plane is mapped into the circumference on
the $z$-plane defined by the equation $|z-1|=1$, while the values
$r=0,\infty$ correspond to the point $z=0$.

The function $w(z)$ is bounded near the point $z=0$ and
$1-\widetilde{\gamma}=-\sqrt{(n-2)^{2}+32a}\leqslant 0$; therefore
it holds (see (\ref{Fseries}))
\begin{align*}
w(z)&=w_{+}(z):=c_{+}F(\widetilde{\alpha},\widetilde{\beta};\widetilde{\gamma};z),\,z\in
\left(|z-1|=1,\,\im z>0\right),\,c_{+}=\const,
\\ w(z)&=w_{-}(z):=c_{-}
F(\widetilde{\alpha},\widetilde{\beta};\widetilde{\gamma};z),\,z\in\left(|z-1|=1,\,\im
z<0\right),\,c_{-}=\const.
\end{align*}
An equivalent problem for the hypergeometric equation was
considered in \cite{Ste}. However there was made an assumption
equivalent to $c_{+}=c_{-}$ without any proof (formula (3) in
\cite{Ste}). Below we fill this gap.

Functions $w_{\pm}(z)$ should be analytic continuations of each
other through the regular point $z=2$.\footnote{Recall that the
function $F(\alpha',\beta';\gamma';z)$ is holomorphic in
$\mathbb{C}\backslash[1,+\infty)$.} Due to formula
(\ref{HypgeomExpansions1}) (applicable since
$\widetilde{\gamma}-\widetilde{\alpha}-\widetilde{\beta}\notin
\mathbb{R}$) it means that functions
$$
c_{+}\frac{\Gamma(\widetilde{\gamma})\Gamma(\widetilde{\gamma}-
\widetilde{\alpha}-\widetilde{\beta})}
{\Gamma(\widetilde{\gamma}-\widetilde{\alpha})
\Gamma(\widetilde{\gamma}-\widetilde{\beta})}F(\widetilde{\alpha},\widetilde{\beta};
\widetilde{\alpha}+\widetilde{\beta}-\widetilde{\gamma}+1;1-z),\,|z-1|=1,\,\im
z>0
$$
and
$$
c_{-}\frac{\Gamma(\widetilde{\gamma})\Gamma(\widetilde{\gamma}-\widetilde{\alpha}-
\widetilde{\beta})}{\Gamma(\widetilde{\gamma}-\widetilde{\alpha})
\Gamma(\widetilde{\gamma}-\widetilde{\beta})}F(\widetilde{\alpha},\widetilde{\beta};
\widetilde{\alpha}+\widetilde{\beta}-\widetilde{\gamma}+1;1-z),\,|z-1|=1,\,\im
z<0
$$
are analytic continuations of each other through the point $z=2$
as well as functions
$$
c_{+}\frac{\Gamma(\widetilde{\gamma})\Gamma(\widetilde{\alpha}+\widetilde{\beta}
-\widetilde{\gamma})}{\Gamma(\widetilde{\alpha})
\Gamma(\widetilde{\beta})}(1-z)^{\widetilde{\gamma}-\widetilde{\alpha}-\widetilde{\beta}}
F(\widetilde{\gamma}-\widetilde{\alpha},
\widetilde{\gamma}-\widetilde{\beta};
\widetilde{\gamma}-\widetilde{\alpha}-\widetilde{\beta}+1;1-z),\,|z-1|=1,\,\im
z>0
$$
and
$$
c_{-}\frac{\Gamma(\widetilde{\gamma})\Gamma(\widetilde{\alpha}+\widetilde{\beta}-
\widetilde{\gamma})}
{\Gamma(\widetilde{\alpha})\Gamma(\widetilde{\beta})}(1-z)^{\widetilde{\gamma}-
\widetilde{\alpha}-\widetilde{\beta}}
F(\widetilde{\gamma}-\widetilde{\alpha},\widetilde{\gamma}-\widetilde{\beta};
\widetilde{\gamma}-\widetilde{\alpha}-\widetilde{\beta}+1;1-z),\,|z-1|=1,\,\im
z<0.
$$
The first requirement is equivalent to the equality
\begin{equation}\label{Lin1}
\left(c_{+}-c_{-}\right)\frac{\Gamma(\widetilde{\gamma})
\Gamma(\widetilde{\gamma}-\widetilde{\alpha}-\widetilde{\beta})}
{\Gamma(\widetilde{\gamma}-\widetilde{\alpha})
\Gamma(\widetilde{\gamma}-\widetilde{\beta})}=0,
\end{equation}
while the second one to the equality
\begin{equation}\label{Lin2}
\left(c_{+}-c_{-}\exp\left(2\pi\ii(\widetilde{\gamma}-\widetilde{\alpha}-
\widetilde{\beta})\right)\right)\frac{\Gamma(\widetilde{\gamma})
\Gamma(\widetilde{\alpha}+\widetilde{\beta}-\widetilde{\gamma})}
{\Gamma(\widetilde{\alpha})\Gamma(\widetilde{\beta})}=0.
\end{equation}
Since
$\widetilde{\gamma}-\widetilde{\alpha}-\widetilde{\beta}\notin\mathbb{R}$
linear system (\ref{Lin1}), (\ref{Lin2}) has a nontrivial solution
$c_{+},c_{-}$ iff either
$$
\frac{\Gamma(\widetilde{\gamma})\Gamma(\widetilde{\gamma}-\widetilde{\alpha}-
\widetilde{\beta})}{\Gamma(\widetilde{\gamma}-\widetilde{\alpha})
\Gamma(\widetilde{\gamma}-\widetilde{\beta})}=0\quad
\text{or}\quad \frac{\Gamma(\widetilde{\gamma})
\Gamma(\widetilde{\alpha}+\widetilde{\beta}-\widetilde{\gamma})}
{\Gamma(\widetilde{\alpha})\Gamma(\widetilde{\beta})}=0.
$$
Taking into account
$\widetilde{\gamma}-\widetilde{\beta}\not\in\mathbb{R},\;\widetilde{\beta}
\not\in\mathbb{R}$, one gets
$\widetilde{\gamma}-\widetilde{\alpha}=-k+1$ or
$\widetilde{\alpha}=-k+1,\,k\in\mathbb{N}$.

Not loosing generality suppose that $\RE s<0$. Then the first
equality is impossible and the second one yields
$$
s=1-2k-\sqrt{(n-2)^{2}+32a}+\frac{4\ii
mR\gamma}{1-2k-\sqrt{(n-2)^{2}+32a}},
$$
since $\im s^{2}=8\ii mR\gamma$. From the definition of $s$ one
gets therefore the following formula for energy levels:
\begin{align*}
E_{k}&=\frac1{mR^{2}}\left(\frac12(k^{2}-k+1)-\frac{n}4
+2a+b+\frac{2k-1}4\sqrt{(n-2)^{2}+32a}\right)\\
&-\frac{2m\gamma^{2}}{\left(\sqrt{(n-2)^{2}+32a}+2k-1\right)^{2}},\,k\in\mathbb{N}.
\end{align*}

These energy levels are degenerated and their multiplicities
coincide with multiplicities of eigenvalues in propositions
\ref{eigenvectorsOdd}, \ref{eigenvectorsk=1} and
\ref{eigenvectorsEven}.

Taking into account all transformations used while reducing
equation (\ref{SpectralEqGenForm}) to the hypergeometric one we
get the following expression for radial eigenfunctions (up to an
arbitrary constant nonzero factor)
\begin{equation*}
f_{k}(r)=\frac{r^{\rho^{(0)}_{+}}(r-\ii)^{\rho^{(\ii)}_{+}}}
{(r+\ii)^{\rho^{(0)}_{+}+\rho^{(\ii)}_{+}+\rho^{(\infty)}_{+}}}
\sum_{j=0}^{k-1}\frac{(-1)^{j}}{j!(k-j-1)!}
\frac{(\widetilde{\beta})_{j}}{(\widetilde{\gamma})_{j}}
\frac{(4r\ii)^{j}}{(r+\ii)^{2j}},
\end{equation*}
where
$\rho^{(0)}_{+},\,\rho^{(\ii)}_{+},\,\rho^{(\infty)}_{+},\widetilde{\beta}$
and $\widetilde{\gamma}$ are given by above formulas for
$E=E_{k}$.

\subsection{Oscillator potential}

The oscillator potential for the sphere ${\bf S}^{n}$ has the form
$$
V_{o}(r)=\frac12R^{2}\omega^{2}\tan^{2}\frac{\rho}R=\frac{2R^{2}\omega^{2}r^{2}}{(1-r^{2})^{2}},\,\omega\in\mathbb{R}_{+}.
$$
It has a positive singularity along the sphere equator and looks
like an infinite potential well. Therefore from the physical point
of view it is natural to consider wave functions defined on $M'$
and vanishing as $r\to 1$.

From the mathematical point of view theorem \ref{MilatTheorem} is
not applicable since
$$V_{o}\not\in\mathcal{L}^{1}_{\loc}\left({\bf S}^{n}\times{\bf
S}^{n},\chi\times\chi\right).$$ However since $V_{o}\geqslant0$
one can use the Friedrichs extension $\left(H_{V_{o}}\right)_{F}$
of a Hamiltonian with the domain given by theorem
\ref{HFDomainTh}, where $M'\subset{\bf S}^{n}\times{\bf S}^{n}$ is
defined by the inequality $r=\tan\left(\rho/(2R)\right)<1$.

Equation (\ref{SpectralEqGenForm}) for $V=V_{o}$ is a Fuchsian one
with six singular points $0,\pm 1,\pm\ii,\infty$ and corresponding
characteristic exponents:
\begin{align*}
&\rho^{(0)}_{\pm}=\frac12\left(2-n\pm\sqrt{(n-2)^{2}+32a}\right),\,
\rho^{(\infty)}_{\pm}=\frac12\left(2-n\pm\sqrt{(n-2)^{2}+32c}\right),\\
&\rho^{(\ii)}_{\pm}=\rho^{(-\ii)}_{\pm}=\frac12(n-1)\pm\frac12
\sqrt{\left(n-1\right)^{2}+
8mER^{2}+4mR^{4}\omega^{2}+8(a-b+c)},\\
&\rho^{(1)}_{\pm}=\rho^{(-1)}_{\pm}=\frac12\left(1\pm\sqrt{1+4R^{4}m\omega^{2}}\right).
\end{align*}

Similarly to the previous section the function $f(r),\; r\in
(0,1)$ should be $\sim r^{\rho^{(0)}_{+}}$ as $r\rightarrow+0$. On
the other hand the inclusion
$$
f(r)\psi_{D}\in W^{1,2}\left(M',\chi\times\chi\right)
$$
implies the convergence of the integral
\begin{equation}\label{IntCong}
\int_{M'}g_{2}\left(\nabla\overline{\left(f\psi_{D}\right)},
\nabla\left(f\psi_{D}\right)\right)d\chi\times d\chi
\end{equation}
where $g_{2}$ is defined in (\ref{g2metric}) and $\nabla$ means
the gradient operator.

The convergence of (\ref{IntCong}) is equivalent to the
convergence of its "radial part"
$$
\int_{0}^{1}|f'(r)|^{2}\frac{r^{n-1}dr}{(1+r^{2})^{n-4}}.
$$
Therefore if $f\sim r^{\rho^{(1)}}$ as $r\rightarrow1-0$, then
$\rho^{(1)}>1/2$ and thus $\rho^{(1)}=r^{\rho^{(1)}_{+}}$.

Conversely, it can be easily verified that if $f$ is a solution of
(\ref{SpectralEqGenForm}) with asymptotics $f(r)\sim
r^{\rho^{(0)}_{+}}$ as $r\rightarrow0$ and $f(r)\sim
r^{\rho^{(1)}_{+}}$ as $r\rightarrow 1-0$ for $V=V_{o}$, then
$f(r)\psi_{D}\in\Dom\left(\left(H_{V_{o}}\right)_{F}\right)$.

Fortunately, one can glue points $r=\pm1$ together (as well as
points $r=\pm\ii$) by the change of the independent variable
$r\to\zeta,\,\zeta=r^{2}$, which transforms the differential
equation under consideration into the following Fuchsian
differential equation with four singular points:
\begin{align}\begin{split}
f_{\zeta\zeta}+\frac{n+(4-n)\zeta}{2\zeta(\zeta+1)}f_{\zeta}+
\frac{2}{\zeta(\zeta+1)^{2}}&\left(mR^{2}\left(E-\frac{2R^{2}\omega^{2}\zeta}{(\zeta-1)^{2}}
\right)-\frac{a}\zeta-b-c\zeta\right)f=0,\\ 0&<\zeta<1.
\end{split}\label{OscilEqZeta}\end{align}
Singular points $-1,0,1,\infty$ of this equation form a harmonic
quadruple and correspond respectively to characteristic exponents:
$$
\rho^{(\ii)}_{\pm},\,\frac12\rho^{(0)}_{\pm},\,
\rho^{(1)}_{\pm},\,\frac12\rho^{(\infty)}_{\pm}.
$$
The same arguments as for the Coulomb problem leads to the
conclusion that the only possibility to transform equation
(\ref{OscilEqZeta}) to the hypergeometric one via transformations
(\ref{Mobius}), (\ref{Homotopy}) and then using theorem
\ref{HeunRedTh} corresponds to the map of the unordered pair
$(0,\infty)$ into the unordered pair $(0,2)$ by a M\"{o}bius
transformation.

Not loosing generality, one can consider the substitution
\begin{equation}\label{ZetaT}
t=\tau(\zeta)=\frac{2\zeta}{\zeta+1},\quad\tau:\;(-1,0,1,\infty)\rightarrow(\infty,0,1,2).
\end{equation}
The interval under consideration for the variable $t$ is again
$(0,1)$. Substitution (\ref{ZetaT}) transforms equation
(\ref{OscilEqZeta}) into equation (\ref{PreHeun}) with
\begin{gather*}
\mathcal{A}(t)=\frac{n(t-1)}{t(t-2)},\;
\mathcal{B}(t)=\frac{2}{t(t-2)}\left(mR^{2}\left(E+\frac{R^{2}\omega^{2}t(t-2)}{2(t-1)^{2}}
\right)-\frac{2a}t+a-b+\frac{ct}{t-2}\right).
\end{gather*}
Define a function $w(t)$ by
$$
w(t)=t^{-\frac12\rho^{(0)}_{+}}(t-1)^{-\rho^{(1)}_{+}}(t-2)^{-\frac12\rho^{(\infty)}_{+}}f(t).
$$
It satisfies Heun equation (\ref{HeunEq}), where
\begin{gather*}
\alpha=\frac12\rho^{(0)}_{+}+\rho^{(1)}_{+}
+\frac12\rho^{(\infty)}_{+}+\rho^{(\ii)}_{+},\,\beta=\frac12\rho^{(0)}_{+}+\rho^{(1)}_{+}
+\frac12\rho^{(\infty)}_{+}+\rho^{(\ii)}_{-},\,d=2,\\
\gamma=1+\frac12\left(\rho^{(0)}_{+}-\rho^{(0)}_{-}\right),\,\delta=1+\rho^{(1)}_{+}
-\rho^{(1)}_{-},\,
\varepsilon=1+\frac12\left(\rho^{(\infty)}_{+}-\rho^{(\infty)}_{-}\right).
\end{gather*}
Here
$t^{-\frac12\rho^{(0)}_{+}}(t-1)^{-\rho^{(1)}_{+}}(t-2)^{-\frac12\rho^{(\infty)}_{+}}$
means the function holomorphic on the domain \\
$\mathbb{C}\backslash\left((-\infty,0]\cup[1,+\infty)\right)$ and
real for real $t\in(0,1)$. Restrictions on asymptotics of the
function $f(r)$ near the points $r=0,1$ are equivalent to the
boundedness of the function $w(t)$ near the points $t=0,1$.

Calculation, similar to (\ref{qCalcul}), yields the following
value of accessory parameter $q$ for (\ref{HeunEq})
$$
q=-2mR^{2}E+2b+n\left(\rho^{(1)}_{+}+\frac14\rho^{(\infty)}_{+}\right)+
2\rho^{(0)}_{+}\rho^{(1)}_{+}+\frac12\rho^{(0)}_{+}\rho^{(\infty)}_{+}+
\frac{n}4\rho^{(0)}_{+}.
$$
Condition (\ref{Gamma=Epsilon}) of theorem \ref{HeunRedTh} is
again equivalent to (\ref{FirstCondit}). Condition
(\ref{q=AlphaBeta}) of the same theorem can be written as
$$
\alpha\beta-q=\rho^{(1)}_{+}\left(\rho^{(\infty)}_{+}-\rho^{(0)}_{+}\right)=0,
$$
which is again a consequence of (\ref{Gamma=Epsilon}).

Suppose that condition (\ref{Gamma=Epsilon}) is valid. Thus we are
in the situation of the first case of theorem \ref{HeunRedTh} and
changing the independent variable $t$ by a new one $z$ according
to (\ref{zt}), one gets hypergeometric equation (\ref{HypEqTilde})
with
\begin{align*}
\widetilde{\alpha}&=\frac12\alpha=\frac14\left(2+\sqrt{(n-2)^{2}+32a}+
\sqrt{1+4R^{4}m\omega^{2}}+s\right),\\
\widetilde{\beta}&=\frac12\beta=\frac14\left(2+\sqrt{(n-2)^{2}+32a}+
\sqrt{1+4R^{4}m\omega^{2}}-s\right),\\
\widetilde{\gamma}&=\gamma=1+\frac12\sqrt{(n-2)^{2}+32a},
\end{align*}
where $s=\sqrt{(n-1)^{2}+8mER^{2}+4mR^{4}\omega^{2}+16a-8b}$. The
interval $(0,1)\ni t$ corresponds to the interval $(0,1)\ni z$,
therefore the requirement on asymptotic of the function $f(t)$
near the point $t=0$ implies
$$
w(z)=F(\widetilde{\alpha},\widetilde{\beta};\widetilde{\gamma};z).
$$
Also due to
$$
\widetilde{\gamma}-\widetilde{\alpha}-\widetilde{\beta}=-\frac12\sqrt{1+4R^{4}m\omega^{2}}<0,\;
\RE\widetilde{\alpha}>0
$$
and (\ref{LimitNear1}), the requirement on asymptotic of the
function $f(t)$ near the point $t=1$ implies
$$
\widetilde{\beta}=-k,\;k=0,1,2,\ldots
$$
This leads to energy levels
\begin{align*}\begin{split}
E_{k}&=\frac1{8mR^{2}}\left(\left(4k+2+\sqrt{(n-2)^{2}+32a}\right)^{2}-(n-1)^{2}-16a+8b+1\right)\\&+
\frac{\omega}{2\sqrt{m}}\left(4k+2+\sqrt{(n-2)^{2}+32a}\right)\sqrt{1+\frac1{4R^{4}m^{2}}},\,k=0,1,2,\ldots
\end{split}
\end{align*}

Again multiplicities of these energy levels coincide with
multiplicities of eigenvalues in propositions
\ref{eigenvectorsOdd}, \ref{eigenvectorsk=1} and
\ref{eigenvectorsEven}.

The expression for radial eigenfunctions (up to an arbitrary
constant nonzero factor) is
\begin{equation*}
f_{k}(r)=\frac{r^{\rho^{(0)}_{+}}(r^{2}-1)^{\rho^{(1)}_{+}}}
{(r^{2}+1)^{\frac12\rho^{(0)}_{+}+\rho^{(1)}_{+}+\frac12\rho^{(\infty)}_{+}}}
\sum_{j=0}^{k}\frac{(-1)^{j}}{j!(k-j)!}
\frac{(\widetilde{\alpha})_{j}}{(\widetilde{\gamma})_{j}}
\frac{4^{j}r^{2j}}{(r^{2}+1)^{2j}},
\end{equation*}
where
$\rho^{(0)}_{+},\,\rho^{(1)}_{+},\,\rho^{(\infty)}_{+},\widetilde{\alpha}$
and $\widetilde{\gamma}$ are given by above formulas for
$E=E_{k}$.

\section{Conclusion}\label{Conclud}\markright{\ref{Conclud} Conclusion}

The possibility to find in explicit way some (but not all)
eigenvalues for a Schr\"{o}dinger operator characterizes so called
quasi exactly solvable models \cite{Us1}--\cite{Ush}. In the
present paper we have shown that the two-body problem on spheres
${\bf S}^{n}$ with Coulomb and oscillator potentials is quasi
exactly solvable for any $n$. A possible generalization for other
compact two-point homogeneous spaces is an open problem.

The quasi exactly solvability here is an attribute not of a radial
differential equation (\ref{SpectralEqGenForm}), but of the whole
problem. It stems from the two causes. The first cause follows
from the fact that we restrict our consideration on the subspace
of $\mathcal{L}(G,K,\mu)$ (see (\ref{SpaceFactorization}))
consisting of common eigenfunctions for operators $D_{0}^{2},
D_{1},D_{2}$ and optionally $D_{3}$. For every such eigenfunction
one gets a separate radial differential equation
(\ref{SpectralEqGenForm}). For the Coulomb and oscillator
potentials this equation can be reduced to the Heun one, but the
further reduction to the hypergeometric equation using Maier's
scheme is possible only for some eigenfunctions (just for those
that satisfies the equation (\ref{FirstCondit})) and this is the
second cause.

\numberwithin{theore}{section} \numberwithin{equation}{section}
\numberwithin{proposit}{section}
\appendix
\section{Orthogonal complex Lie algebras and their
representations}\label{Appendix A}\markright{\ref{Appendix A}
Orthogonal complex Lie algebras and their representations}
\subsection{Lie algebra $\mathfrak{B}_{k}$}\label{BCaseAp}

Here is a brief description of the simple complex Lie algebra
$\mathfrak{B}_{k}\cong\mathfrak{so}({2k+1,\mathbb{C}})$ (see
\cite{Hamphreys}, \cite{GotoGross} and \cite{VO} for details).

Denote
$$
S_{i}=\begin{pmatrix}0 & 0 & \dots & 0 & 0 & 1 \\ 0 & 0 & \dots &
0 & 1 & 0 \\ \vdots & \vdots & \ddots & \vdots & \vdots \\ 1 & 0 &
\ldots & 0 & 0 & 0
\end{pmatrix}\in \GL(i),\,i\in\mathbb{N}.
$$
Consider the Lie algebra
$\mathfrak{B}_{k}\cong\mathfrak{so}({2k+1,\mathbb{C}})$ as
\begin{equation}\label{BkForm}
\mathfrak{B}_{k}=\left(\left.A\in\mathfrak{gl}(2k+1,\mathbb{C})\right|\;A^{T}S_{2k+1}+
S_{2k+1}A=0\right).
\end{equation}
Following \cite{Mol2} we shall enumerate the rows and columns of
$A\in\mathfrak{B}_{k}$ by the indices $-k,\ldots,-1,0,1,\ldots,k$.
The convenience of such notations is due to the fact that
subalgebras $\mathfrak{B}_{i}\subset\mathfrak{B}_{k},\, i<k$
correspond to indices of rows and columns from $-i$ to $i$.

It can be easily shown that a matrix
$$
A=\sum_{i,j}a_{ij}E_{ij}\in\mathfrak{gl}(2k+1,\mathbb{C})
$$
belongs to $\mathfrak{B}_{k}$ iff $a_{ij}+a_{-j,-i}=0$, which
means that $A$ is skew-symmetric w.r.t.\ its secondary diagonal.

Let $ F_{ij}=E_{ij}-E_{-j,-i}$. It is easily seen that
$$
[F_{ij},F_{pq}]=\delta_{jp}F_{iq}-\delta_{iq}F_{pj}+
\delta_{-pi}F_{-qj}+\delta_{-jq}F_{p,-i}.
$$
The algebra $\mathfrak{B}_{k}$ is spanned by elements $F_{ij}$
with $i>-j$. Evidently, $F_{i,-i}=0$ and $F_{-j,-i}=-F_{ij}$.

Elements $F_{ii},\,i=1,\ldots,k$ form a base of the Cartan
subalgebra $\mathfrak{h}_{k}\subset\mathfrak{B}_{k}$, which
consists of elements of the form
$$
X=\diag\left(-x_{k},-x_{k-1},\ldots,-x_{1},0,x_{1},\ldots,x_{k-1},x_{k}\right).
$$
Let $\varepsilon_{i}\in\mathfrak{h}_{k}^{*}$ such that
$\varepsilon_{i}(X)=x_{i}$, i.e.\ $\varepsilon_{i}$ is a base in
$\mathfrak{h}_{k}^{*}$ dual to $F_{i,i},\,i=1,\ldots,k$. Define a
symmetric nondegenerate bilinear form $\langle\cdot,\cdot\rangle$
on $\mathfrak{B}_{k}$ as
\begin{equation}\label{TraceForm}
\langle A,B\rangle=\frac12\tr AB,
\end{equation}
which is proportional to the Killing form. Clearly,
$$
\langle F_{ij},F_{qp}\rangle=\delta_{ip}\delta_{jq},\,i>-j,q>-p.
$$
In particular, $F_{ii},\,i=1,\ldots,k$ is an orthogonal base in
$\mathfrak{h}_{k}$.

The form
$\left.\langle\cdot,\cdot\rangle\right|_{\mathfrak{h}_{k}}$
generates the isomorphism
$\varkappa:\;\mathfrak{h}_{k}\to\mathfrak{h}_{k}^{*}$ by the
formula $\varkappa(X)=\langle X,\cdot\rangle$.  Specifically,
$\varkappa(F_{i,i})=\varepsilon_{i}$ and
$\varepsilon_{i},\,i=1,\ldots,k$ is an orthonormal base in
$\mathfrak{h}_{k}^{*}$ w.r.t.\ the form
$$\langle f_{1},f_{2}\rangle^{*}:=
\langle\varkappa^{-1}(f_{1}),\varkappa^{-1}(f_{2})\rangle,\,f_{1},f_{2}
\in\mathfrak{h}_{k}^{*}.$$

Using this notation one can describe the standard form of the root
system for $\mathfrak{B}_{k}$ in the following way. Let
$$
\Phi_{\mathfrak{B}_{k}}:=\left(\left.\pm\varepsilon_{i},\pm\varepsilon_{i}
\pm\varepsilon_{j}\right|\;i\ne j,\,i,j=1,\ldots,k\right)
$$
be a root system in $\mathfrak{B}_{k}$,
$$
\Phi_{\mathfrak{B}_{k}}^{+}:=\left(\left.\varepsilon_{i},\varepsilon_{i}
+\varepsilon_{j},\varepsilon_{i}-\varepsilon_{j}\right|\;i>j,\,i,j=1,\ldots,k\right)
$$
be a system of positive roots, and
$$
\Delta_{\mathfrak{B}_{k}}:=\left(\left.\alpha_{1}=\varepsilon_{1},\alpha_{i}=\varepsilon_{i}-
\varepsilon_{i-1}\right|\,i=2,\ldots,k\right)
$$
be a system of simple roots, corresponding to the inverse
lexicographic order. A subalgebra
$\mathfrak{B}_{i}\subset\mathfrak{B}_{k},\,i<k$ corresponds to
root systems
$\Phi_{\mathfrak{B}_{i}},\,\Phi_{\mathfrak{B}_{i}}^{+}$ and
$\Delta_{\mathfrak{B}_{i}}$.

Let $L_{\alpha}$ be a root subspace in $\mathfrak{B}_{k}$,
corresponding to a root $\alpha\in\Phi_{\mathfrak{B}_{k}}$. Then
\begin{gather}\begin{split}
L_{-\varepsilon_{i}}=\lspan(F_{0i}),\,L_{\varepsilon_{i}}=\lspan(F_{i0}),\,
L_{\varepsilon_{i}-\varepsilon_{j}}=\lspan(F_{ij}),\,\\
L_{\varepsilon_{i}+\varepsilon_{j}}=\lspan(F_{i,-j}),\,
L_{-\varepsilon_{i}-\varepsilon_{j}}=\lspan(F_{-ij}),\,i,j=1,\ldots,k.
\end{split}\label{WeigtSubspacesB}\end{gather}

Fundamental weights for $\mathfrak{B}_{k}$ are
$$
\lambda_{1}=\frac12\sum_{j=1}^{k}\varepsilon_{j},\,
\lambda_{i}=\sum_{j=i}^{k}\varepsilon_{j},\;i=2,\ldots,k.
$$
Let
$$
\lambda=\sum_{j=1}^{k}\lambda^{j}\lambda_{j},\,\lambda^{j}\in\mathbb{Z}_{+}:=(0)\cup\mathbb{N}
$$
be a dominant weight and $V(\lambda)$ be an irreducible finite
dimensional $\mathfrak{B}_{k}$-module with the highest weight
$\lambda$. All finite dimensional irreducible representations of
$\mathfrak{B}_{k}$ are of this form, modules $V(\lambda)$ with
different $\lambda$ are not isomorphic to each other, and
$V(\lambda)$ corresponds to a (single valued) representation of
the group $\SO(2k+1)$ iff $\lambda_{1}$ is even. The dominant
weight $\lambda$ can be written in the form
\begin{equation}\label{HigestWeightB}
\lambda=\sum_{i=1}^{k}m_{i}\varepsilon_{i},\,m_{k}\geqslant
m_{k-1}\geqslant\ldots \geqslant m_{1}\geqslant0,
\end{equation}
where either all $m_{i}\in\mathbb{Z}_{+}$ or all
$m_{i}\in\mathbb{Z}_{+}+\frac12$. Even values of $\lambda_{1}$
corresponds to $m_{i}\in\mathbb{Z}_{+}$. Let $\delta$ be the sum
of fundamental weights. Then it holds
\begin{equation}\label{DeltaDefB}
\delta=\sum_{i=1}^{k}\lambda_{i}=\frac12\sum_{\alpha\in\Phi_{\mathfrak{B}_{k}}^{+}}^{k}\alpha=
\sum_{i=1}^{k}\left(i-\frac12\right)\varepsilon_{i}.
\end{equation}
The universal Casimir operator $C\in U(\mathfrak{B}_{k})$ is
\begin{equation}\label{Casimir}
C=\sum_{i=1}^{k}\left(F_{ii}^{2}+\{F_{i0},F_{0i}\}\right)+
\sum_{i>j>0}\left(\{F_{ij},F_{ji}\}+\{F_{i,-j},F_{-ji}\}\right).
\end{equation}
The following formulas are valid for any semisimple Lie algebra:
\begin{gather}\label{CasimirEigen}
\left.C\right|_{V(\lambda)}=\left(\langle\delta+\lambda,\delta+\lambda\rangle-
\langle\delta,\delta\rangle\right)\id,\\ \label{Weyl} \dim
V(\lambda)=\left.\prod_{\alpha\succ
0}\langle\lambda+\delta,\alpha\rangle\right/\prod_{\alpha\succ
0}\langle\delta,\alpha\rangle,
\end{gather}
where $\alpha\succ 0$ means a positive root.

For any semisimple Lie algebra $\mathfrak{g}$ and its Cartan
subalgebra $\mathfrak{h}$ the module $V(\lambda)$ can be
decomposed into the finite direct sum of weight subspaces
$$
V(\lambda)=\bigoplus_{\mu}V_{\mu}(\lambda),\,\mu\in\mathfrak{h}^{*},
$$
where $\forall v\in V_{\mu}(\lambda),\forall h\in\mathfrak{h}$ it
holds $h(v)=\mu(h)v$ and the sum is over weights of the form
$$
\lambda-\sum_{\alpha\succ0}i_{\alpha}\alpha,\,i_{\alpha}\in\mathbb{Z}_{+}.
$$
Besides, for any root $\alpha$ of $\mathfrak{g}$ one has
\begin{equation}\label{WeightMove}
\xi_{\alpha}:\,V_{\mu}(\lambda)\rightarrow
V_{\mu+\alpha}(\lambda),\;\xi_{\alpha}\in L_{\alpha}.
\end{equation}

\subsection{Lie algebra $\mathfrak{D}_{k}$}\label{DCaseAp}

The Lie algebra $\mathfrak{D}_{k}$ is the subalgebra of
$\mathfrak{B}_{k}$, consisting of matrices whose column and rows
with the index $0$ vanish. We shall discard these null row and
column and shall enumerate other rows and columns of
$A\in\mathfrak{D}_{k}$ by the indices $-k,\ldots,-1,1,\ldots,k$ as
before. The Cartan subalgebra
$\mathfrak{h}_{k}\subset\mathfrak{D}_{k}$ is the same as in the
$\mathfrak{B}_{k}$-case. Describe the $\mathfrak{D}_{k}$-case
briefly, emphasizing differences from the $\mathfrak{B}_{k}$-case.

Now one has
\begin{gather*}
\Phi_{\mathfrak{D}_{k}}:=\left(\left.\pm\varepsilon_{i}
\pm\varepsilon_{j}\right|\;i\ne j,\,i,j=1,\ldots,k\right),\\
\Phi_{\mathfrak{D}_{k}}^{+}:=\left(\left.\varepsilon_{i}
+\varepsilon_{j},\varepsilon_{i}-\varepsilon_{j}\right|\;i>j,\,i,j=1,\ldots,k\right),\\
\Delta_{\mathfrak{D}_{k}}:=\left(\left.\alpha_{1}=\varepsilon_{1}+\varepsilon_{2},
\alpha_{i}=\varepsilon_{i}-\varepsilon_{i-1}\right|\,i=2,\ldots,k\right).
\end{gather*}
The root subspaces $L_{\pm\varepsilon_{i}\pm\varepsilon_{j}}$ are
the same as in $\mathfrak{B}_{k}$-case.

Fundamental weights are
$$
\lambda_{1}=\frac12\sum_{j=1}^{k}\varepsilon_{j},\,
\lambda_{2}=-\frac12\varepsilon_{1}+\frac12\sum_{j=2}^{k}\varepsilon_{j},\,
\lambda_{i}=\sum_{j=i}^{k}\varepsilon_{j},\;i=3,\ldots,k.
$$
The sum of fundamental weights is
$$
\delta=\sum_{i=1}^{k}\lambda_{i}=\frac12\sum_{\alpha\in\Phi_{\mathfrak{D}_{k}}^{+}}^{k}\alpha=
\sum_{i=2}^{k}(i-1)\varepsilon_{i}.
$$

A dominant weight
$$
\lambda=\sum_{j=1}^{k}\lambda^{j}\lambda_{j},\,\lambda^{j}\in\mathbb{Z}_{+}:=(0)\cup\mathbb{N}
$$
now has the form
\begin{equation}\label{HigestWeightD}
\lambda=\sum_{i=1}^{k}m_{i}\varepsilon_{i},\,m_{k}\geqslant
m_{k-1}\geqslant\ldots\geqslant m_{2}\geqslant |m_{1}|,
\end{equation}
where either $m_1\in\mathbb{Z},
m_{i}\in\mathbb{Z}_{+},\,i\geqslant2$ or
$m_1\in\mathbb{Z}+\frac12,
m_{i}\in\mathbb{Z}_{+}+\frac12,\,i\geqslant2$. Again
$\mathfrak{D}_{k}$-modules with integer $m_j,\,j=1,\ldots,k$
correspond to (single valued) representations of the group
$\SO(2k)$.

The universal Casimir operator $C\in U(\mathfrak{D}_{k})$ is
\begin{equation}\label{CasimirD}
C=\sum_{i=1}^{k}F_{ii}^{2}+
\sum_{i>j>0}\left(\{F_{ij},F_{ji}\}+\{F_{i,-j},F_{-ji}\}\right).
\end{equation}

\subsection{Restrictions of $\mathfrak{B}_{k}$ and
$\mathfrak{D}_{k}$-representations.}

The following results were found in \cite{Zh1962} (see also
\cite{Zhelob}).

Let $V_{\mathfrak{B}_{k}}(\lambda)$ be a simple
$\mathfrak{B}_{k}$-module with a highest weight
(\ref{HigestWeightB}) and $V_{\mathfrak{D}_{k}}(\lambda)$ be a
simple $\mathfrak{D}_{k}$-module with a highest weight
\begin{equation*}
\lambda'=\sum_{i=1}^{k}m'_{i}\varepsilon_{i},\,m'_{k}\geqslant
m'_{k-1}\geqslant\ldots\geqslant m'_{2}\geqslant |m'_{1}|.
\end{equation*}
\begin{proposit}\label{BtoDRestriction}
The restriction
$\left.V_{\mathfrak{B}_{k}}(\lambda)\right|_{\mathfrak{D}_{k}}$ of
the irreducible $\mathfrak{B}_{k}$-representation onto any
subalgebra $\mathfrak{D}_{k}\subset\mathfrak{B}_{k}$ expands as
follows
$$
\left.V_{\mathfrak{B}_{k}}(\lambda)\right|_{\mathfrak{D}_{k}}=
\bigoplus_{\lambda'}V_{\mathfrak{D}_{k}}(\lambda'),
$$
where the summation is over all $\lambda'$ such that
$$m_{k}\geqslant m'_{k}\geqslant m_{k-1}\geqslant\ldots\geqslant
m'_{2}\geqslant m_{1}\geqslant m'_{1}\geqslant -m_{1}$$ and all
$m'_j$ are integer or half integer simultaneously with $m_j$.
\end{proposit}

Let $V_{\mathfrak{B}_{k-1}}(\lambda')$ be a simple
$\mathfrak{B}_{k-1}$-module with a highest weight
\begin{equation*}
\lambda'=\sum_{i=1}^{k-1}m'_{i}\varepsilon_{i},\,m'_{k-1}\geqslant
m'_{k-2}\geqslant\ldots\geqslant m'_{2}\geqslant m'_{1}\geqslant
0.
\end{equation*}

\begin{proposit}\label{DtoBRestriction}
The restriction
$\left.V_{\mathfrak{D}_{k}}(\lambda)\right|_{\mathfrak{B}_{k-1}}$
of the irreducible $\mathfrak{D}_{k}$-representation onto any
subalgebra $\mathfrak{B}_{k-1}\subset\mathfrak{D}_{k}$ expands as
follows
$$
\left.V_{\mathfrak{D}_{k}}(\lambda)\right|_{\mathfrak{B}_{k-1}}=\bigoplus_{\lambda'}V_{\mathfrak{B}_{k-1}}(\lambda')
$$
where the summation is over all $\lambda'$ such that
$$m_{k}\geqslant m'_{k-1}\geqslant m_{k-1}\geqslant\ldots\geqslant
m_{2}\geqslant m'_{1}\geqslant |m_{1}|$$ and all $m'_j$ are
integer or half integer simultaneously with $m_j$.
\end{proposit}

\section{Self-adjointness of Schr\"{o}dinger operators on Riemannian spaces}\label{AppendixE}
\markboth{\ref{AppendixE} Self-adjointness of Schr\"{o}dinger
operators}{\ref{AppendixE} Self-adjointness of Schr\"{o}dinger
operators}

Here we shall formulate two results concerning the
self-adjointness of Schr\"{o}dinger operators on Riemannian
spaces, which is used in section \ref{SpectralEquationTBP}.

The first theorem is a result from \cite{Mil1}, restricted onto
the scalar case.
\begin{theore}\label{MilatTheorem}
Let $M$ be a Riemannian manifold of a bounded geometry, $\dim
M=\ell$, and $\mu$ be the measure on $M$ generated by its metric.
Suppose also that the potential $V$ can be represented in the form
$V=V_{1}+V_{2}$, where real valued functions $V_{1},V_{2}$ are as
follows: $0\leqslant
V_{1}\in\mathcal{L}^{1}_{\loc}(M,\mu),\;0\geqslant
V_{2}\in\mathcal{L}^{p}(M,\mu)$ for $p=\ell/2$ if
$\ell\geqslant3$, for $p>1$ if $\ell=2$, and for $p=1$ if
$\ell=1$.

Then the operator $H_{V}=-\laplace+V$ is self-adjoint with the
domain:
\begin{equation}\label{MilatDomain}
\Dom(H_{V})=\left(\left.u\in W^{1,2}(M,\mu)\right|\;
\int_{M}V_{1}|u|^{2}d\mu<+\infty,\;H_{V}u\in\mathcal{L}^{2}(M,\mu)\right),
\end{equation}
where $H_{V}u$ is understood in the sense of distributions. Here
$W^{1,2}(M,\mu)$ is the Sobolev space, consisting of functions on
$M$ that are in $\mathcal{L}^{2}(M,\mu)$ with their first
derivatives.

Also $Vu\in\mathcal{L}^{1}_{\loc}(M,\mu)$ for $u\in\Dom(H_{V})$.
\end{theore}

The definition of a Riemannian manifold of a bounded geometry can
be found in \cite{Shub}. Note that compact and homogeneous
Riemannian manifold is always of a bounded geometry.

If the potential $V$ is not in $\mathcal{L}^{1}_{\loc}(M^{n},\mu)$
then theorem \ref{MilatTheorem} is not applicable. If instead $V$
is bounded from below, one can try to restrict the Schr\"{o}dinger
operator onto some submanifold $M'$ of $M^{\ell}$ such that
$\left.V\right|_{M'}\in\mathcal{L}^{1}_{\loc}(M',\mu)$ and
construct the Friedrichs self-adjoin extension \cite{SR2} of
$-\laplace+V$ from the initial domain $C^{\infty}_{c}(M')$. This
procedure is physically motivated for instance in the case when
$V\rightarrow+\infty$ near the boundary of $M'$ and therefore wave
functions should vanish near this boundary.

Let us turn to the accurate mathematical description. Let $M'$ be
an open connected submanifold of a Riemannian space $M^{\ell}$ of
dimension $\ell$ with a metric $g$ and an induced measure $\mu$.
We do not suppose that $M'$ is complete w.r.t.\ the Riemannian
structure induced by the Riemannian structure on $M^{\ell}$. Let
$V\geqslant C\in\mathbb{R}$ be a real valued function from
$\mathcal{L}^{1}_{\loc}(M',\mu)$ and $H'=-\laplace+V$ be a
Schr\"{o}dinger operator with the domain $C^{\infty}_{c}(M')$,
consisting of all infinitely smooth complex valued functions in
$M'$ with compact supports. Not loosing generality we suppose that
$C=1$. Let $H_{F}\geqslant\id$ be the abstract Friedrichs
extension of $H'$ \cite{SR2}. We need a precise description of
$\Dom(H_{F})$.

The operator $H'$ generates sesquilinear nonnegative form $q_{H'}$
by the equality
\begin{equation*}\label{qH'}
q_{H'}(\varphi,\psi)=\int_{M'}\left(\overline{H'\varphi}\right)\psi
d\mu
\end{equation*}
with the domain $C^{\infty}_{c}(M')$. Evidently, its closure is
\begin{equation}\label{qClosureOnManifold}
q_{H_{F}}(\varphi,\psi)=\int_{M'}\left(g(\nabla\bar\varphi,\nabla\psi)+
V\bar\varphi\psi\right)d\mu
\end{equation} with
$\Dom(q_{H_{F}})\subset\mathcal{L}^{2}(M',\mu)$ being a closure of
$C^{\infty}_{c}(M')$ w.r.t.\ the inner product
(\ref{qClosureOnManifold}), where $\nabla$ is the gradient
operator given in local coordinates by the equality
$$
\nabla\psi=g^{jk}\frac{\partial \psi}{\partial
x^{k}}\frac{\partial}{\partial x^{j}}.
$$

The operator $H_{F}$ is defined by the identity
$$
\int_{M'}\left(g(\nabla\bar\varphi,\nabla\psi)+V\bar\varphi\psi\right)d\mu=
\int_{M'}\bar\varphi H_{F}\psi
d\mu,\,\forall\varphi\in\Dom(q_{H_{F}}),\psi\in\Dom(H_{F}).
$$
Thus
\begin{equation}\label{HFdist}
H_{F}\psi=\left(-\laplace\psi+V\psi\right)_{\dist},\,\psi\in\Dom(H_{F}).
\end{equation}
\begin{theore}\label{HFDomainTh}
The domain of the operator $H_{F}$ is
\begin{equation*}\label{HFDomain}
\left(\left.\psi\in
W^{1,2}\left(M',\mu\right)\,\right|\,V\psi\in\mathcal{L}^{1}_{\loc}(M',\mu);
\left(-\laplace\psi+V\psi\right)_{\dist}\in\mathcal{L}^{2}(M',\mu)\right)
\end{equation*}
and $H_{F}$ acts by formula {\rm(\ref{HFdist})}.
\end{theore}
The proof of this theorem repeats {\it mutatis mutandis} the proof
of theorem X.27 from \cite{SR2} using the generalization of the
{\it Kato inequality} for Riemannian spaces \cite{Br}.

\section{Some Fuchsian differential
equations}\label{AppendixD} \markboth{\ref{AppendixD} Some
Fuchsian differential equations}{\ref{AppendixD} Some Fuchsian
differential equations}

For convenience of references we collected here basic facts
concerning some Fuchsian differential equations: the Riemannian
equation and the reducibility of the Heun equation to the
hypergeometric one.

The linear differential equation
\begin{equation}\label{FuchsEquation}
w^{(n)}(z)+p_{1}(z)w^{(n-1)}(z)+\ldots+p_{n}(z)w(z)=0
\end{equation}
on the Riemannian sphere
$\overline{\mathbb{C}}=\mathbf{P}^{1}(\mathbb{C})$ with
meromorphic coefficients $p_{i}(z),\,i=1,\ldots,n$ is {\it
Fuchsian} if for any $z_{0}\in\overline{\mathbb{C}}$ its solutions
has no more than a power growth as $z$ tends to $z_{0}$ in some
cone\footnote{In a neighborhood of the infinite point one should
use the local coordinate $\zeta=1/z$ instead of $z-z_{0}$.}, not
containing a neighborhood of its vertex $z_{0}$. A point $z_{0}$
is {\it regular} for this differential equation if all solutions
of (\ref{FuchsEquation}) are holomorphic in some neighborhood of
$z_{0}$; otherwise $z_{0}$ is a {\it singular point}.

It is known \cite{CodLev} that equation (\ref{FuchsEquation}) is
Fuchsian one iff
$$
p_{i}(z)=\frac{q_{i}(z)}{\prod\limits_{k=1}^{m}(z-z_{k})^{i}}
$$
for some finite potentially singular points
$z_{1},\ldots,z_{m}\in\mathbb{C}$ and polynomials $q_{i}(z)$ of
degrees $\leqslant i(m-1)$. One can find {\it characteristic
exponents} $\rho^{(z_{k})}$ of (\ref{FuchsEquation}) at the point
$z_{k}$ by the substitution $w(z)=(z-z_{k})^{\rho^{(z_{k})}}$ into
(\ref{FuchsEquation}) and keeping only leading terms as
$z\rightarrow z_{k}$. This procedure gives an algebraic equation
of the $n$-th degree for $\rho^{(z_{k})}$. Denote by
$\rho^{(z_{k})}_{i},\,i=1,\ldots,n$ its solutions for all points
$z_{k},\,k=1,\ldots,m$. The substitution
$w(z)=z^{-\rho^{(\infty)}}$ similarly gives characteristic
exponents $\rho^{(\infty)}_{1},\ldots,\rho^{(\infty)}_{n}$ in the
infinity. Characteristic exponents satisfy the {\it Fuchs
identity\label{FuchsIdentity}}:
$$
\sum_{i=1}^{n}\sum_{k=1}^{m+1}\rho^{(z_{k})}_{i}=\frac12(m-1)n(n-1),
$$
where $\rho^{(z_{m+1})}_{i}:=\rho^{(\infty)}_{i}$.

One can find characteristic exponents also for a regular point. If
a point $\tilde z$ is regular, then characteristic exponents for
this point are $0,1,\ldots,n-1$. The sufficient condition for the
regularity of $\tilde z$ is the regularity of coefficients
$p_{i}(z),\,i=1,\ldots,n$ at this point.

An information on singular points and corresponding characteristic
exponents of equation (\ref{FuchsEquation}) can be encoded in the
{\it Riemann $P$-symbol\label{RiemannSymbol}}
$P\{\mathcal{A};z\}$, where the first row of a matrix
$\mathcal{A}$ consists of singular points and other rows of
$\mathcal{A}$ consist of corresponding characteristic exponents.

Equation (\ref{FuchsEquation}) of the second order with three
singular points is called the {\it Riemannian
equation\label{RiemannEquation}}. Coefficients of the Riemann
equation are completely defined by its characteristic exponents.
Equivalently, the Riemann equation is completely defined by its
$P$-symbol. In this case the Fuchs identity looks like
$$
\sum_{i=1}^{2}\sum_{k=1}^{3}\rho^{(z_{k})}_{i}=1
$$
and there are only five independent characteristic values.

There are two types of variable change, transforming any Fuchsian
equation into another Fuchsian equation. The first one is a {\it
linear-fractional (M\"{o}bius)
transformation\label{LinFracMobTransf}} of the independent
variable:
\begin{equation}\label{Mobius}
z\rightarrow t,\;z=\frac{\alpha t+\beta}{\gamma
t+\delta},\,\alpha\delta-\beta\gamma\ne0.
\end{equation}
By such transformation one can move three singular points into
three arbitrary points of $\overline{\mathbb{C}}$ with the same
characteristic exponents.

The second one is a linear transformation of the dependent
variable
\begin{equation}\label{Homotopy}
w(z)\rightarrow
w_{1}(z)=\left(\frac{z-z_{1}}{z-z_{2}}\right)^{q}w(z),
\end{equation}
which conserves singular points, but changes the characteristic
exponents
$$
\rho^{(z_{1})}_{i}\rightarrow\rho^{(z_{1})}_{i}+q,\,\rho^{(z_{2})}_{i}\rightarrow\rho^{(z_{2})}_{i}-q,\,i=1,2.
$$

Using these transformation for the Riemannian equation one can
move three singular points into the triple $(0,1,\infty)$ such
that $\rho^{(0)}_{1}=\rho^{(1)}_{1}=0$. If one denote
$\rho^{(\infty)}_{1}=\alpha,\rho^{(\infty)}_{2}=\beta$ and
$\rho^{(0)}_{2}=1-\gamma$, then the Fuchs identity for this
equation gives $\rho^{(1)}_{2}=\gamma-\alpha-\beta$ that
corresponds to the {\it
hypergeometric\label{HypergeometricEquation}} or {\it Gauss
equation}:
\begin{equation}\label{HypEquation}
z(1-z)w''(z)+(\gamma-(\alpha+\beta+1)z)w'(z)-\alpha\beta w(z)=0.
\end{equation}
The $P$-symbol of equation (\ref{HypEquation}) is
$$
P\left\{\begin{matrix} 0 & 1 & \infty & \\ 0 & 0 & \alpha & ;z \\
1-\gamma & \gamma-\alpha-\beta & \beta &
\end{matrix}\right\}.
$$

Many quantum mechanical problems for constant curvature spaces can
be reduced to this equation, while their Euclidean counterparts
lead to its limiting cases, obtained from (\ref{HypEquation}) by
confluence of singular points (such equations are not Fuchsian).

We shall consider only solutions of (\ref{HypEquation}) in the
case $\gamma\ne-m,\,m\in\mathbb{N}$. Solutions of
(\ref{FuchsEquation}), corresponding to different characteristic
exponents near some singular point are called {\it canonical
solutions\label{CanonicalSolution}} near that point. The series
\begin{equation}\label{Fseries}
F(\alpha,\beta;\gamma;z):=\sum_{n=0}^{\infty}\frac{(\alpha)_{n}(\beta)_{n}}{(\gamma)_{n}}\frac{z^{n}}{n!},\;
|z|<1
\end{equation}
where $(a)_{n}:=a(a+1)\ldots(a+n-1),\,(a)_{0}:=1$, is the
canonical solution of (\ref{HypEquation}), corresponding to the
characteristic exponent $\rho^{(0)}_{1}=0$. The function
$F(\alpha,\beta;\gamma;z)$, defined by (\ref{Fseries}) for $|z|<1$
can be analytically continued for
$z\in\mathbb{C}\backslash(1,+\infty)$ \cite{Go}, \cite{Abra}.

Evidently $F(\alpha,\beta;\gamma;z)=F(\beta,\alpha;\gamma;z)$. If
$\alpha=-m$ or $\beta=-m,\,m=0,1,2,\ldots$, then
$F(\alpha,\beta;\gamma;z)$ is a polynomial of degree $m$.

Another canonical solution of (\ref{HypEquation}), corresponding
to the characteristic exponent $\rho^{(0)}_{2}=1-\gamma$ for
$\gamma\notin\mathbb{N}$, is
$$z^{1-\gamma}F(\alpha-\gamma+1,\beta-\gamma+1;2-\gamma;z).$$
Canonical solutions near the singular point $z=1$ are
$$
F(\alpha,\beta;\alpha+\beta-\gamma+1;1-z)$$ and if
$\gamma-\alpha-\beta\notin\mathbb{Z}$ also
$$
(1-z)^{\gamma-\alpha-\beta}F(\gamma-\alpha,\gamma-\beta;\gamma-\alpha-\beta+1;1-z).
$$
Near the singular point $z=\infty$ canonical solutions are
$$
z^{-\alpha}F(\alpha,\alpha-\gamma+1;\alpha-\beta+1;\frac1z),\;z^{-\beta}F(\beta,\beta-\gamma+1;\beta-\alpha+1;\frac1z)
$$
if $\alpha-\beta\notin\mathbb{Z}$. If $\alpha-\beta\in\mathbb{Z}$,
then only one of these expressions is a canonical solution: the
first if $\alpha-\beta>0$ or the second if $\alpha-\beta<0$.

There are expansions of $F(\alpha,\beta;\gamma;z)$ through
canonical solutions near the singular points $z=1$ and $z=\infty$
\cite{Abra}, \cite{BeEr}, important for spectral problems. The
first one, used in the present paper, is
\begin{align}\label{HypgeomExpansions1}
&F(\alpha,\beta;\gamma;z)=\frac{\Gamma(\gamma)\Gamma(\gamma-\alpha-\beta)}{\Gamma(\gamma-\alpha)\Gamma(\gamma-\beta)}
F(\alpha,\beta;\alpha+\beta-\gamma+1;1-z)\\&+
\frac{\Gamma(\gamma)\Gamma(\alpha+\beta-\gamma)}{\Gamma(\alpha)\Gamma(\beta)}
(1-z)^{\gamma-\alpha-\beta}F(\gamma-\alpha,\gamma-\beta,\gamma-\alpha-\beta+1,1-z),\,
|\arg(1-z)|<\pi\notag
\end{align}
if $\gamma-\alpha-\beta\notin\mathbb{Z}$. Here $\Gamma$ is the
{\it gamma-function}. It has no zeros and has poles of the first
order at the points $z=-m,\,m=0,1,2,\ldots$. Its logarithmic
derivative $\psi_{\Gamma}(z):=\Gamma'(z)/\Gamma(z)$ also has poles
of the first order at the same points.

For $\gamma-\alpha-\beta\in\mathbb{Z}$ every summand at the right
hand side of (\ref{HypgeomExpansions1}) is singular and it holds
for $m=0,1,2,\ldots$
\begin{align}\label{HypgeomExpansions11}
F(\alpha,&\beta;\alpha+\beta+m;z)=\frac{\Gamma(m)\Gamma(\alpha+\beta+m)}{\Gamma(\alpha+m)\Gamma(\beta+m)}
\sum\limits_{n=0}^{m-1}\frac{(\alpha)_{n}(\beta)_{n}}{n!(1-m)_{n}}(1-z)^{n}\notag\\&-
\frac{\Gamma(\alpha+\beta+m)}{\Gamma(\alpha)\Gamma(\beta)}(z-1)^{m}
\sum\limits_{n=0}^{\infty}\frac{(\alpha+m)_{n}(\beta+m)_{n}}{n!(n+m)!}(1-z)^{n}
\\ \label{HypgeomExpansions12}&\times(\ln(1-z)-\psi_{\Gamma}(n+1)
-\psi_{\Gamma}(n+m+1)+\psi_{\Gamma}(\alpha+n+m)+\psi_{\Gamma}(\beta+n+m))\notag,\\
F(\alpha,&\beta;\alpha+\beta-m;z)=\frac{\Gamma(m)\Gamma(\alpha+\beta-m)}{\Gamma(\alpha)\Gamma(\beta)}
(1-z)^{-m}\sum\limits_{n=0}^{m-1}\frac{(\alpha-m)_{n}(\beta-m)_{n}}{n!(1-m)_{n}}(1-z)^{n}\notag\\&-
\frac{(-1)^{m}\Gamma(\alpha+\beta-m)}{\Gamma(\alpha-m)\Gamma(\beta-m)}
\sum\limits_{n=0}^{\infty}\frac{(\alpha)_{n}(\beta)_{n}}{n!(n+m)!}(1-z)^{n}
\\&\times(\ln(1-z)-\psi_{\Gamma}(n+1)
-\psi_{\Gamma}(n+m+1)+\psi_{\Gamma}(\alpha+n)+\psi_{\Gamma}(\beta+n))\notag,\\
&|\arg(1-z)|<\pi,\;|1-z|<1\notag.
\end{align}
In the case $\RE(\gamma-\alpha-\beta)<0$ formulas
(\ref{HypgeomExpansions1}) -- (\ref{HypgeomExpansions12}) imply
\begin{equation}\label{LimitNear1}
\lim_{z\rightarrow1}F(\alpha,\beta;\gamma;z)(1-z)^{\alpha+\beta-\gamma}=
\frac{\Gamma(\gamma)\Gamma(\alpha+\beta-\gamma)}{\Gamma(\alpha)\Gamma(\beta)}.
\end{equation}

The Fuchsian equation (\ref{FuchsEquation}) with four singular
points by transformations (\ref{Mobius}) and (\ref{Homotopy}) can
be reduced to the {\it Heun equation\label{HeunEquation}}
\begin{equation}\label{HeunEq}
w''(t)+\left(\frac{\gamma}t+\frac{\delta}{t-1}+\frac{\varepsilon}{t-d}\right)w'(t)+
\frac{\alpha\beta t-q}{t(t-1)(t-d)}w(t)=0,
\end{equation}
where $0,1,d,\infty$ are its four singular points ($d\neq
0,1,\infty$) and $\alpha+\beta-\gamma-\delta-\varepsilon+1=0$. The
corresponding $P$-symbol is
$$
P\left\{\begin{matrix} 0 & 1 & d & \infty & \\ 0 & 0 & 0 & \alpha
&;t\\ 1-\gamma & 1-\delta & 1-\varepsilon & \beta &
\end{matrix}\right\}.
$$
Note that the {\it accessory parameter} $q$ does not arise in this
$P$-symbol.

The theory of the Heun equation is much less explicit than the
theory of the Riemannian equation. In particular, there are no
explicit expressions of canonical solutions near different
singular points through each other. Therefore there are only
approximate methods for solving spectral problems connected with
the Heun equation, using continued fractions (see for example
\cite{SlavLay} and references therein).

The substitution $z=P(t)$ for a rational function $P$ transforms
equation (\ref{FuchsEquation}) into another Fuchsian equation with
generally a greater number of singular points. Therefore sometimes
the inverse transformation can decrease the number of singular
points of a Fuchsian equation.\footnote{Generally, the inverse
transformation does not conserve the Fuchs class of differential
equations.}

At the present time there is no a general theory of such
reduction. However in \cite{Maier} there were classified all Heun
equations (\ref{HeunEq}) that can be obtained by a substitution
$z=P(t)$ from the hypergeometric one (\ref{HypEquation}). By the
inverse transformation these Heun equations are reduced to
hypergeometric equations.

The first condition for existing such reduction is the position of
the point $d$. Let
$$
(z_{1},z_{2},z_{3},z_{4})_{c.r.}:=\frac{(z_{1}-z_{3})(z_{2}-z_{4})}{(z_{1}-z_{4})(z_{2}-z_{3})}
$$
be the {\it cross-ratio} of four pairwise distinct points from
$\overline{\mathbb{C}}$. It is well known that a cross-ration is
invariant under M\"{o}bius transformations. The group
$\mathfrak{S}_{4}$, permuting points $z_{1},z_{2},z_{3}$ and
$z_{4}$, acts on their cross-ration. The cross-ration orbit
$\mathcal{O}_{\mathfrak{S}_{4}}(s)$ of $s:=(z_{1},z_{2},z_{3},\\
z_{4})_{c.r.}$ consists of points
$s,1-s,1/s,1/(1-s),s/(s-1),(s-1)/s\in\overline{\mathbb{C}}$.

In general position this orbit consists of six points, but there
are two exceptional cases: the orbit $-1,\frac12,2$ and the orbit
$\frac12\pm\frac{\sqrt{3}}2\ii$. If
$(z_{1},z_{2},z_{3},z_{4})_{c.r.}\in (-1,\frac12,2)$, then
$(z_{1},z_{2},z_{3},z_{4})$ is a {\it harmonic
quadruple\label{HarmonicQuadruple}}. If
$(z_{1},z_{2},z_{3},z_{4})_{c.r.}=\frac12\pm\frac{\sqrt{3}}2\ii$,
then $(z_{1},z_{2},z_{3},z_{4})$ is an {\it equianharmonic
quadruple\label{EquianharmonicQuadruple}}.

Points of a harmonic quadruple lie on a circle or on a line. By a
M\"{o}bius transformation they can be mapped into vertices of a
square in $\mathbb{C}$. If $(z_{1},z_{2},z_{3},\infty)$ is a
harmonic quadruple, then $(z_{1},z_{2},z_{3})$ are collinear,
equally spaced points. If $(z_{1},z_{2},z_{3},\infty)$ is an
equianharmonic quadruple, then $(z_{1},z_{2},z_{3})$ are vertices
of an equilateral triangle in $\mathbb{C}$.

\begin{theore}[\cite{Maier}]\label{HeunRedTh}
All cases, when nontrivial Heun equation {\rm(\ref{HeunEq})}
(i.e.\ $\alpha\beta\neq 0$ or $q\neq 0$) can be obtained from the
hypergeometric one {\rm(\ref{HypEquation})} by the rational
substitution $z=P(t)$, are as follows.
\begin{enumerate}
\item Harmonic case: $d\in\mathcal{O}_{\mathfrak{S}_{4}}(2)$.
Suppose $d=2$,\footnote{If
$d\in\mathcal{O}_{\mathfrak{S}_{4}}(s)$, then the quadruple
$(0,1,d,\infty)$ can be mapped into the quadruple $(0,1,s,\infty)$
by a M\"{o}bius transformation, which transforms also parameters
of equation (\ref{HeunEq}).} then $q/(\alpha\beta)$ must be equal
$1$, and characteristic exponents of points $t=0$ and $t=d=2$ must
be the same, i.e.\ $\gamma=\varepsilon$. The function $P(t)$ is a
degree-$2$ polynomial and can be chosen as
$P(t)=t(2-t)=1-(t-1)^{2}$. It maps $t=0,2$ to $z=0$ and $t=1$ to
$z=1$.\footnote{This transformation was found already in
\cite{Kuiken}.}

If additionally $1-\delta=2(1-\gamma)$, then $P(t)$ can be chosen
also as degree-$4$ polynomial $4(t(2-t)-\frac12)^{2}$, which maps
$t=0,1,2$ to $z=1$.

\item $d\in\mathcal{O}_{\mathfrak{S}_{4}}(4)$.
Suppose $d=4$, then $q/(\alpha\beta)$ must be equal $1$,
characteristic exponents of points $t=1$ must be double those of
the point $t=d=4$, i.e.\ $1-\delta=2(1-\varepsilon)$, and $t=0$
must have characteristic exponents $0,1/2$, i.e.\
$\gamma=\frac12$. The function $P(t)$ is a degree-$3$ polynomial
and can be chosen as $(t-1)^{2}(1-\frac{t}4)$. It maps $t=0$ to
$z=1$ and $t=1,4$ to $z=0$.

\item Equianharmonic case: $d\in\mathcal{O}_{\mathfrak{S}_{4}}(\frac12+\frac{\sqrt{3}}2\ii)$.
Characteristic exponents of points $t=0,1,d$ are the same, i.e.\
$\gamma=\delta=\varepsilon$. Suppose
$d=\frac12+\frac{\sqrt{3}}2\ii$, then $q/(\alpha\beta)$ must be
equal $\frac12+\frac{\sqrt{3}}6\ii$. The function $P(t)$ is a
degree-$3$ polynomial and can be chosen as
$\left(1-t/(\frac12+\frac{\sqrt{3}}6\ii)\right)^{3}$. It maps
$t=0,1,d$ to $z=1$ and $t=q/(\alpha\beta)$ to $z=0$, thus creating
a new singular point.

If additionally $\gamma=\delta=\varepsilon=\frac23$, then $P(t)$
can be chosen also as degree-$6$ polynomial
$$4\left(\left(1-\frac{t}{\frac12+\frac{\sqrt{3}}6\ii}\right)^{3}-\frac12\right)^{2},$$
which maps $t=0,1,d,q/(\alpha\beta)$ to $z=1$.

\item $d\in\mathcal{O}_{\mathfrak{S}_{4}}(\frac12+\frac{5\sqrt{2}}4\ii)$.
Suppose $d=\frac12+\frac{5\sqrt{2}}4\ii$, then $q/(\alpha\beta)$
must be equal $\frac12+\frac{\sqrt{2}}4\ii$, characteristic
exponents of the point $t=d$ must be $0,1/3$, i.e.\
$\varepsilon=2/3$, and points $t=0,1$ must have characteristic
exponents $0,1/2$, i.e.\ $\gamma=\delta=1/2$. The function $P(t)$
is a degree-$4$ polynomial and can be chosen as
$$\left(1-\frac{t}{\frac12+\frac{5\sqrt{2}}4\ii}\right)
\left(1-\frac{t}{\frac12+\frac{\sqrt{2}}4\ii}\right)^{3}.$$ It
maps $t=0,1$ to $z=1$ and $t=d,q/(\alpha\beta)$ to $z=0$.

\item $d\in\mathcal{O}_{\mathfrak{S}_{4}}(\frac12+\frac{11\sqrt{15}}{90}\ii)$.
Suppose $d=\frac12+\frac{11\sqrt{15}}{90}\ii$, then
$q/(\alpha\beta)$ must be equal $\frac12+\frac{\sqrt{15}}{18}\ii$,
characteristic exponents of the point $t=d$ must be $0,1/2$, i.e.\
$\varepsilon=1/2$, and points $t=0,1$ must have characteristic
exponents $0,1/3$, i.e.\ $\gamma=\delta=2/3$. The function $P(t)$
is a degree-$5$ polynomial and can be chosen as
$$-\ii\frac{2025\sqrt{15}}{64}t(t-1)
\left(t-\frac12-\frac{\sqrt{15}}{18}\ii\right)^{3}.$$ It maps
$t=0,1,q/(\alpha\beta)$ to $z=0$ and $t=d$ to $z=1$.
\end{enumerate}

\end{theore}

Note that there are three independent parameters in the first case
of theorem (\ref{HeunRedTh}) (for example: $\alpha,\beta,\gamma$)
and all other cases contain only one or two free parameters. It
means that the first case is more rife in applications. In fact,
it is the only one, which occurs in the present paper.

\section{The proof of two expansions}\label{appD}
\markboth{\ref{appD} The proof of two expansions}{\ref{appD} The
proof of two expansions}

Expansions (\ref{MainExpansion}), (\ref{MainExpansion1}) were
obtained in \cite{Mol2} and \cite{Mol1} using the theory of
Yangians, which is a part of the quantum algebra. Here we give an
independent proof of these expansions from a classical point of
view using one result from \cite{Di}. The initial idea of this
proof is due to A.I.~Molev.

Let $\mathcal{A}$ be an associative algebra over $\mathbb{C}$,
generated by elements $Z_{+},Z_{-},F$ and relations
\begin{align}\label{UnCommute}
[F,Z_{+}]=2Z_{+},\,[F,Z_{-}]=-2Z_{-},\,[Z_{+},Z_{-}]=-\frac12F^{3}+qF,\,q\in\mathbb{C}.
\end{align}
Let also $\tau:\,\mathcal{A}\to\Hom_{\mathbb{C}}(V)$ be its
irreducible linear representation in a finite dimensional complex
linear space $V$. An arbitrary linear operator in $V$ has at least
one eigenvalue. Let
$F\chi=\eta\chi,\,\eta\in\mathbb{\mathbb{C}},\,\chi\in
V,\chi\neq0$, then relations (\ref{UnCommute}) imply
$FZ_{+}\chi=(\eta+2)Z_{+}\chi,\,FZ_{-}\chi=(\eta-2)Z_{-}\chi$.
Since $\dim_{C}V<\infty$ there is a vector $v_{\nu}\in
V,\,v_{\nu}\neq0$ such that
$Fv_{\nu}=\nu,\,Z_{+}v_{\nu}=0,\,\nu\in\mathbb{C}$. Let
$v_{\nu-2j}:=Z^{j}_{-}v_{\nu},\,j\in\mathbb{Z}_{+}$, then
\begin{equation}\label{2Rel}
Fv_{\eta}=\eta
v_{\eta},\,Z_{-}v_{\eta}=v_{\eta-2},\,\forall\eta\in
L_{\nu}:=\nu-2\mathbb{Z}_{+}.
\end{equation}
Let $\mu$ be a root of equation
\begin{equation}\label{MuEq}
\mu^{2}+2\mu+\nu^{2}+2\nu-4q=0.
\end{equation}
\begin{Lem}\label{lemD}
It holds
\begin{equation}\label{3Rel}
Z_{+}v_{\eta}=\frac1{16}(\eta-\mu)(\eta-\nu)(\eta+\mu+2)(\eta+\nu+2)v_{\eta+2},\,
\eta\in L_{\nu}.
\end{equation}
\end{Lem}
\begin{proof}
For $\eta=\nu$ equality (\ref{3Rel}) is obvious. Let $\alpha\in
L_{\nu},\,\alpha\leqslant\nu$ and suppose that (\ref{3Rel}) is
valid for any $\eta\in L_{\nu}$ such that $\eta>\alpha$. Then one
gets
\begin{align*}
Z_{+}&v_{\alpha}=Z_{+}Z_{-}v_{\alpha+2}=[Z_{+},Z_{-}]v_{\alpha+2}+Z_{-}Z_{+}v_{\alpha+2}=
\left(-\frac12F^{3}+qF\right)v_{\alpha+2}\\&+\frac1{16}(\alpha+2-\mu)(\alpha+2-\nu)(\alpha+\mu+4)
(\alpha+\nu+4)Z_{-}v_{\alpha+4}\\&=\left(-\frac12(\alpha+2)^{3}+q(\alpha+2)+
\frac1{16}(\alpha+2-\mu)(\alpha+2-\nu)(\alpha+\mu+4)(\alpha+\nu+4)\right)v_{\alpha+2}\\&=
\frac1{16}(\alpha-\mu)(\alpha-\nu)(\alpha+\mu+2)(\alpha+\nu+2)v_{\alpha+2},
\end{align*}
due to (\ref{UnCommute}), (\ref{2Rel}), (\ref{3Rel}) and the
following identity
\begin{gather*}
(\alpha+2-\mu)(\alpha+2-\nu)(\alpha+\mu+4)(\alpha+\nu+4)-
(\alpha-\mu)(\alpha-\nu)(\alpha+\mu+2)(\alpha+\nu+2)\\=8(\alpha+2)^{3}-16q(\alpha+2).
\end{gather*}
This completes the induction.
\end{proof}
Due to (\ref{2Rel}) nonzero vectors $v_{\eta},\,\eta\in L_{\nu}$
are linear independent, therefore from $\dim_{\mathbb{C}}V<\infty$
one gets $v_{\nu_{1}}\neq0,\,v_{\nu_{1}-2}=0$ for some $\nu_{1}\in
L_{\nu}$. Since the representation $\tau$ is irreducible one gets
\begin{equation}\label{Vspan}
V=\lspan\left(v_{\nu_{1}},v_{\nu_{1}+2},\ldots,v_{\nu}\right),\,\dim_{\mathbb{C}}V
=\frac12(\nu-\nu_{1})+1
\end{equation}
due to (\ref{2Rel}) and (\ref{3Rel}). Equation (\ref{3Rel}) for
$\eta=\nu_{1}-2$ implies
\begin{equation}\label{conclusion}
(\nu_{1}-2-\mu)(\nu_{1}+\mu)(\nu_{1}+\nu)=0.
\end{equation}

Let now $Z_{+}=D^{+},Z_{-}=D^{-},F=F_{kk},\,k\geqslant2$ for
operators $D^{+},D^{-},F_{kk}$ from section \ref{n=2k},
$V:=\widetilde{V}_{\mathfrak{B}_{k}}(m_{k}\varepsilon_{k}+m_{k-1}\varepsilon_{k-1})$
and therefore
\begin{equation}\label{q51}
q=\frac12\left(m_{k}^{2}+m_{k-1}^{2}+(2k-1)m_{k}+(2k-3)m_{k-1}\right)+\frac14(2k-1)(2k-3).
\end{equation}
Due to theorem 9.1.12 from \cite{Di} the $\mathcal{A}$-module $V$
is irreducible, therefore lemma \ref{lemD} is applicable. We shall
demonstrate that
\begin{equation}\label{noneq1}
(\nu_{1}-2-\mu)(\nu_{1}+\mu)\neq0.
\end{equation}
Indeed the $\mathcal{A}$-module
$\widetilde{V}_{\mathfrak{B}_{k}}(m_{k}\varepsilon_{k}+m_{k-1}\varepsilon_{k-1})$
does not contain weights $j\varepsilon_{k}$ for $|j|>m_{k}$
\cite{Hamphreys}, therefore
\begin{equation}\label{noneq2}
(\nu_{1}-1)^{2}+(\nu+1)^{2}\leqslant2(m_{k}+1)^{2}.
\end{equation}
If $(\nu_{1}-2-\mu)(\nu_{1}+\mu)=0$ then $|\nu_{1}-1|=|\mu+1|$ and
equation (\ref{MuEq}) leads to the inequality
$$
2+4q=(\mu+1)^{2}+(\nu+1)^{2}\leqslant2(m_{k}+1)^{2},
$$
which is equivalent to
\begin{equation}\label{noneq3}
m_{k-1}^{2}+(2k-3)(m_{k}+m_{k-1})+\frac12(2k-1)(2k-3)\leqslant0
\end{equation}
due to (\ref{q51}). Obviously inequality (\ref{noneq3}) is
impossible.

Thus from (\ref{conclusion}), (\ref{noneq1}) and (\ref{DIM}) one
gets $\nu=-\nu_{1}=m_{k}-m_{k-1}$. Now expansion
(\ref{MainExpansion}) follows from (\ref{2Rel}) and (\ref{Vspan}).

The consideration for the space
$V:=\widetilde{V}_{\mathfrak{D}_{k}}(m_{k}\varepsilon_{k}+m_{k-1}\varepsilon_{k-1}),\,k\geqslant2$
from section \ref{n=2k-1} is similar. Here
$$
q=\frac12\left(m_{k}^{2}+m_{k-1}^{2}+2(k-1)m_{k}+2(k-2)m_{k-1}\right)+(k-1)(k-2)
$$
and the conjecture $(\nu_{1}-2-\mu)(\nu_{1}+\mu)=0$ now implies
due to (\ref{noneq2})
$$
2(k-2)m_{k}+(m_{k-1}+k-2)^{2}+k(k-2)\leqslant0
$$
that leads to $k=2,\,m_{k-1}=0$ and to the equality in
(\ref{noneq2}). This yields
\begin{equation}\label{final}
\nu=-\nu_{1}=m_{k}-|m_{k-1}|.
\end{equation}
The last possibility $\nu_{1}=-\nu$ in (\ref{conclusion}) also
leads to (\ref{final}) due to (\ref{DIM2}) and (\ref{Vspan}). Now
expansion (\ref{MainExpansion1}) is a consequence of (\ref{2Rel}),
(\ref{Vspan}) and (\ref{final}).

\end{document}